\documentclass[manuscript]{acmart}
\AtBeginDocument{%
  }

\setcopyright{acmlicensed}
\copyrightyear{2018}
\acmYear{2018}
\acmDOI{XXXXXXX.XXXXXXX}
\acmConference[Conference acronym 'XX]{Make sure to enter the correct
  conference title from your rights confirmation email}{June 03--05,
  2018}{Woodstock, NY}
\acmISBN{978-1-4503-XXXX-X/2018/06}

\usepackage[table]{xcolor}
\usepackage{longtable}
\usepackage{pdflscape}   
\usepackage{multirow}
\usepackage{booktabs}
\begin{document}

\title{Available but Not Usable: Dark Patterns and Interaction Cost in Social Media Privacy and Safety Settings for Teens}

\author{Jingxin Dong}
\email{dong11@iu.edu}
\orcid{0009-0003-8699-6776}
\affiliation{
  \institution{Indiana University}
  \city{Bloomington}
  \state{Indiana}
  \country{USA}
}

\author{Lingyun Chen}
\email{lch2@iu.edu}
\orcid{0009-0007-9191-0964}
\affiliation{
  \institution{Indiana University}
  \city{Bloomington}
  \state{IN}
  \country{USA}
}

\author{Chen Ling}
\email{ccling@iu.edu}
\orcid{0000-0001-5082-7557}
\affiliation{
  \institution{Indiana University}
  \city{Bloomington}
  \state{Indiana}
  \country{USA}
}

\author{Colin M. Gray}
\email{comgray@iu.edu}
\orcid{0000-0002-7307-1550}
\authornote{Colin has engaged in paid expert witness and consulting work on cases relating to social media. This expert witness work is independent from the data and methods presented in this paper.}
\affiliation{
  \institution{Indiana University}
  \city{Bloomington}
  \state{Indiana}
  \country{USA}
}

\begin{abstract}
Social media platforms are central to teenagers' lives, and their designs can expose users to privacy, safety, and wellbeing harms. Platforms increasingly offer protective settings, though the presence of a control reveals little about whether teenagers can find, use, and benefit from it over time. We paired an expert evaluation of six privacy and safety tasks across TikTok, Instagram, Snapchat, and YouTube with moderated think-aloud sessions in which 11 teenagers aged 14 to 17 attempted the tasks. Interaction cost and dark patterns analysis allowed us to compare the complexity designed into each task with the effort participants incurred as they located, configured, and interpreted controls. Recurring dark patterns appeared across tasks, and most participant attempts exceeded the expert baseline. Protective settings therefore risk being insufficiently usable or durable in practice, and we propose a wayfinding audit that integrates expert evaluation, usability testing, interaction cost, and dark pattern analysis.
\end{abstract}



\received{20 February 2007}
\received[revised]{12 March 2009}
\received[accepted]{5 June 2009}

\maketitle

\section{Introduction}

Social media platforms reach billions of users worldwide and now coordinate much of how teenagers socialize, build friendships, and work out who they are \citep{Anderson2018-qz, Hiniker2018-xz}. Adolescence is also a time when autonomy over personal information begins to matter, and teenagers develop privacy practices through the social relationships a platform mediates \citep{Zhao2022Privacy, Wisniewski2018PrivacyParadox}. Attention capture strategies and other engagement oriented designs therefore reach a population whose developmental position makes the exposure consequential, which has drawn sustained scholarly and regulatory attention. From coordinated actions by 42 attorneys general to a New Mexico judgment against Meta and a Los Angeles County verdict against Meta and Google's YouTube, proceedings across the United States have placed social media design and youth protection under mounting scrutiny \citep{NYAG2023MetaYouth, Ortutay2026MetaLawsuits}. Legislative action has moved in parallel, and Australia now bars users under 16 from the platforms examined here while several US states condition teen accounts on age verification and parental consent \citep{australia2024minimumage, florida2024hb3}.

Platforms have introduced teen accounts and a growing set of privacy, safety, and wellbeing controls over the past several years---created well after the attention capture functionality they are meant to offset. Platforms position privacy and safety settings as tools for managing exposure, regulating use, and responding to unwanted interactions. However, the presence of a setting reveals little about the interaction cost involved in locating it, interpreting its terminology, configuring its options, or determining whether an action has taken effect \citep{Schaffner2022AccountDeletion, Vaccaro2018-yx, Hsu2025-ag}. Some safeguards need to be enabled, some are weakened by less protective defaults, and some offer a temporary alternative that expires without notice. A protective state also has to be durable once a teenager reaches a limit, and research on digital wellbeing tools reports that limits users set are frequently snoozed, deleted, or worked around \citep{mongeroffarello2019race}. Harm to teenagers may therefore continue where safeguards exist and stay too hard to use or too easy to displace to deliver protection. Platform communications establish that a control is available, but scholarship also establishes that manipulative designs recur across privacy and account management \citep{Bosch2016-ei, kelly2023documenting, Schaffner2022AccountDeletion}. Less is known about the interaction cost \citep{Lam2008-fh} teenagers encounter when they attempt concrete privacy and safety tasks on platforms they already use. A control can appear adequate in a feature inventory, and it can seem to be efficient to locate for an expert, while remaining difficult for a teenager to discover and complete \citep{habib2020scavenger, Zhao2022Privacy}. Comparing the interaction cost built into a documented route with the cost that emerges during a teenager's own attempt shows where an available choice falls short of a usable one.

Our study is guided by four research questions:

\begin{description}
    \item[\textbf{RQ1a:}] What interaction cost does each platform require to complete six
    common privacy and safety tasks on TikTok, Instagram, Snapchat, and YouTube, when the
    task goal, entry point, and direct route are known?
    \item[\textbf{RQ1b:}] What dark patterns are present along the routes to the same six
    privacy and safety tasks?
    \item[\textbf{RQ2a:}] What interaction cost do teenagers incur when completing these
    tasks on platforms they already use?
    \item[\textbf{RQ2b:}] Where and how does teen interaction cost diverge from the expert
    baseline?
\end{description}

We conducted a two part study. We first conducted an expert evaluation of six common privacy and safety tasks across TikTok, Instagram, Snapchat, and YouTube, documenting the route to each control, the interface states encountered, the interaction cost it requires, and the dark patterns these paths include. We then conducted moderated think aloud sessions \citep{EricssonSimon1993} with 11 teenagers aged 14 to 17 on platforms they regularly used. Participants shared their phone screens while completing six common privacy and safety tasks. We reconstructed each attempt from the recordings and coded time, clicks, interface states, scrolling, completion, and observable struggle. Setting the two components beside each other makes the cost visible to an evaluator who knows the route comparable with the cost a participant incurs when the route has to be found, interpreted, and completed.

Our contributions are three-fold. First, we operationalize interaction cost of social media settings for comparative analysis, which supplies a vocabulary and a set of measures that separate the cost built into a platform route from the cost that accumulates through search, interpretation, backtracking, and uncertainty. Second, we demonstrate the limits of existing teen safeguards and the prevalence of dark patterns. Most participant attempts cost more than the documented route requires, including costs accrued before participants entered the route, demonstrating the challenges of efficacy. 
Third, we develop an initial wayfinding audit, a technique that brings together dark patterns analysis, temporal analysis, interaction cost measurement, expert evaluation, and usability testing. The audit examines whether a safeguard exists, what its documented route requires, and whether intended users can find it, use it, and read its outcome, which supports diagnosis by designers before deployment and evaluation by researchers, auditors, and regulators retrospectively.

\section{Related Work}
\label{sec:related-work}

\subsection{Social Media, Dark Patterns, and Teenagers}
\label{sec:rw-dark-patterns}

Dark pattern scholarship documents how interface design steers users toward outcomes that serve the platform~\cite{Brignull2010DarkPatterns, mathur2019dark, DiGeronimo2020DarkPatterns}, and social media is increasingly of interest to researchers when considering the power of strategies for capturing attention and governing user choice~\cite{Mildner2023-dg, Seyson2025-dw, Mildner2024-vs}. \citet{Mildner2023Engaging} contributed a foundational account of dark patterns on social media,  providing a thematic analysis of four major social networking services that separates \textit{engaging strategies}, which prolong use, from \textit{governing strategies}, which narrow the choices a user is able to make for themselves. \citet{Monge-Roffarello2023-qc} sharpen our understanding of engagement strategies by treating \textit{attention capture} as a design goal that can be specified and identified through concrete interface features. Privacy and safety controls are where governing strategies become most visible, since a user who wants protection has to work through whatever interface the platform provides~\cite{gray2021dark}. B\"{o}sch et al.\ named privacy dark strategies as a distinct family of manipulative design, and later research also accounted for granularity at the level of individual controls, showing how the strategies structure the privacy choices social networking sites offer~\cite{Bosch2016-ei, kelly2023documenting}. \citet{Schaffner2022AccountDeletion} show what the strategies cost a user in practice, tracing account deletion routes that vary sharply in length and clarity from one platform to the next. Work to consolidate the field has followed~\cite{Gray2023-ay, gray2023dark}, with Gray et al.\ harmonizing ten regulatory and academic taxonomies into a three-level ontology that supplies researchers with a shared vocabulary for naming manipulative design~\cite{gray2024ontology}. A subsequent knowledge stack extends the vocabulary outward, connecting evidence about interface design to form legal and regulatory arguments~\cite{Gray2026-ks, Santos2024-bh}.

Research on children and adolescents extends the same account to users who meet manipulative design as an ordinary condition of platform use~\cite{Rossi2024-dn, Landesman2024-vs, Hardwick2025-bt,Gairola2026-do}. Radesky et al.\ establish the scale of the problem beyond social media proper, identifying manipulative features across mobile applications used by children~\cite{Radesky2022-pc}. Fitton et al.\ organize such features within the 4Cs framework, treating dark patterns as a category of online risk alongside risks involving content and contact~\cite{Fitton2021-ou}. User studies complicate any assumption that awareness alone produces resistance~\cite{DiGeronimo2020DarkPatterns, Sanchez-Chamorro2025-sk}. Sch\"{a}fer et al.\ find that children recognize malicious interface designs and comply with them anyway, and a later study reports adolescents who accept deceptive designs when a platform appears to offer no genuine alternative~\cite{Schafer2024GrowingUp, Schafer2025-xw}. S\'{a}nchez Chamorro et al.\ show how teenagers interpret manipulative designs through advice inherited from parents and peers~\cite{Sanchez-Chamorro2024-hv}, while Kelly and Burkell find that resistance to privacy dark patterns depends on whether a teenager can read a pattern as manipulation in the first place~\cite{Kelly2025-bx}. Chen et al.\ document the designs very large online platforms use to prolong teenage engagement~\cite{Chen2024-hm}, and Natarajan turns from platform design to user response, reporting teenagers who add their own friction to infinite scroll when platform controls do not support disengagement~\cite{Natarajan2024-do, Lukoff2021-yi, Lukoff2023-xz}. We extend the line of work by measuring the interaction cost teenagers incur inside platform-provided privacy and safety controls, and by treating dark patterns as one property of the routes teenagers are made to travel.

\subsection{Usable Privacy and Security}
\label{sec:rw-usable-privacy}

Usable privacy and security research asks whether people can understand and operate the protections systems claim to provide, and the founding studies of the field found that users often cannot. Whitten and Tygar reached the conclusion from a laboratory study of encryption software and Adams and Sasse from organizational password practice, with both reporting that people fail or work around a mechanism whose design ignores what they already know and do~\cite{Whitten1999-jc, Adams1999-un}. Cranor's assessment of standardized notice and choice holds that the mechanism is necessary without being sufficient~\cite{cranor2012necessary}. Schaub et al.\ and Acquisti et al.\ locate the insufficiency in design, the first by showing how rarely a notice offers a real choice at the moment the choice matters, and the second by showing how defaults and framing steer a decision before a person consciously evaluates the options~\cite{Schaub2015-ds, Acquisti2017-nu}.

Later work shifts the focus from the existence of a control toward the effort required to find and use it. Habib et al.\ mapped opt-out and data deletion choices across 150 websites and then tested whether people could complete the tasks, finding mechanisms that were inconsistently labeled and buried several layers below the surface, and participants who gave up on sites that formally offered the option~\cite{Habib2019-dl, habib2020scavenger}. Feng et al.\ and Habib and Cranor turn the finding into apparatus, the first by organizing privacy choices around when a choice appears and what it permits, and the second by proposing criteria a researcher can apply to a choice mechanism consistently across systems~\cite{feng2021designspace, habib2022evaluating}. The field now treats the presence of a control as weak evidence about whether a person can reach it~\cite{habib2020scavenger, Vaccaro2018-yx, Hsu2025-ag}.
 
Social media research asks whether privacy controls remain understandable and usable when their consequences unfold across audiences and over time. Liu et al.\ and Madejski et al.\ established the baseline empirically, the first by comparing what people believed their Facebook settings did against what the settings actually did, and the second by identifying sharing violations that remained invisible to the people affected~\cite{Liu2011-fb, Madejski2012-er}. Mondal et al.\ extend the problem past immediate misconfiguration, arguing that a static setting becomes ineffective as the meaning of shared content changes over time~\cite{Mondal2019-jw}. Measurement work in the privacy and security venues audits what services actually do for younger users, testing platform conduct against what platforms claim~\cite{Reyes2018-cp, Figueira2024-ls}. Recent work turns from user understanding toward the protections platforms choose to provide, with Habib et al.\ documenting a gap between the advertising controls Facebook offers and the forms of control users seek, and Haime and Biddle comparing well-being and moderation tools across several platforms~\cite{Habib2022-ad, Haime2025-ex}. Evaluations of platform-provided protections have generally examined one platform or one mechanism at a time~\cite{Liu2011-fb, Madejski2012-er, Habib2022-ad}. We evaluate privacy and safety controls across four platforms and treat each control as a route with a measurable cost, which lets us ask what a control demands of the person trying to reach it.

\section{Method}

We used a two-part study design to examine the interaction costs participants encountered when completing privacy and safety tasks on four social media platforms. First, we conducted an expert evaluation that documented the structure of each privacy and safety task flow on four platforms, including the minimum clicks and interface states required to reach the intended endpoint. Second, we conducted moderated think-aloud task sessions in which participants (n=11) completed the same tasks on two platforms they reported using regularly. The expert evaluation provides a structural baseline for each platform and task, while the think-aloud sessions show how participants navigated the same controls in practice. We used the two data collection methods together to characterize interaction cost, task completion, and struggle.

\subsection{Part 1: Expert Evaluation of Platform Task Flows}

\subsubsection{Platforms and Tasks}

We evaluated privacy and safety controls across TikTok, Instagram, Snapchat, and YouTube. These features were selected because teenagers use them widely and because they follow different structural and interaction conventions~\cite{Anderson2018-qz, Landesman2024-vs, Chen2024-hm, Hiniker2018-xz}. We examined six tasks that appear commonly in prior research on teen safety, privacy controls, and self-regulation on social media~\cite{Mildner2023-dg, Gruzd2018-zu, Fahlman2018-kw, Schaffner2022AccountDeletion, Zhang2022-bg, Hong2025-pc, Pangrazio2018-zx}, including: setting a screen time limit, making an account private, managing in-app notifications, reporting content, downloading account data, and deleting an account. Not every task was available or applicable on every platform, and we treated each available platform and task combination as a separate task flow.

\subsubsection{Expert Evaluation Procedure}
\label{sec:expert-procedure}
The lead researcher and a senior author with expertise in dark patterns completed the selected tasks in November 2025, using research accounts created with birth dates under 18 so that the accounts reflected each platform's age-based settings.\footnote{TikTok Version 42.2.0; Instagram Version 406.1.0; Snapchat Version 13.59.0; YouTube Version 20.50.9; iOS 26.} For each task, we began from the platform home screen and recorded every interaction required to locate and complete the setting, capturing screenshots of each element of the task flow and organized them sequentially in Miro to reconstruct the flow. For each flow we recorded the minimum number of clicks, the number and type of interface states, the presence of scrolling, the sequence of pages encountered, and any alternative prompts or routes presented before completion. The expert paths served as structural baselines for the participant task attempts, establishing what interaction a task required when the goal, entry point, and direct route were known. The expert paths begin from the platform home screen, whereas participant attempts begin from wherever they concluded the preceding task. 

\subsubsection{Operationalizing Interaction Cost}

We operationalized a \textit{click} as a discrete tap that produced an interface response, and we distinguished among three types of interface states. A \textit{screen} represented a full page transition to a meaningfully different interface state, a \textit{popup} represented a centered modal dialog requiring a response, and a \textit{layover} represented a bottom sheet displayed above the current screen while the underlying interface remained visible. For each task we counted the minimum number of clicks and interface states required for successful completion, and we documented scrollable pages, alternative actions, and design elements that could delay, redirect, or complicate completion. We used the dark patterns ontology from \citet{gray2024ontology} to characterize potentially manipulative or obstructive elements within the flows, coding for the high-level strategies of obstruction, sneaking, interface interference, forced action, and social engineering following prior dark pattern taxonomies ~\cite{Gray2018DarkPatterns, gray2024ontology, Mildner2023-dg}. Two researchers reviewed the flows and resolved disagreements through discussion, and the last author audited the final mappings. The team created analytic notes for uncertain or platform-specific cases.

\subsection{Part 2: Teen Think-Aloud Task Study}

\subsubsection{Participants and Recruitment}

We recruited 11 teenagers between the ages of 14 and 17 who actively used at least two of the four study platforms, reaching them through their parents since participants were minors. We contacted parents through a standing participant registry maintained by one member of the research team, in which parents have previously agreed to be contacted about research involving their children. We emailed registry parents a screening questionnaire that established eligibility and collected background information about their teen's social media use, including which of the four platforms their teen used most often. Parents completed the questionnaire on behalf of their teen, enrolled their teen if interested, and we then scheduled a session with the family. We also used snowball sampling through participating parents' own networks to reach additional teenagers. We used the screening responses to assign each participant two of the four platforms they reported using regularly. Table~\ref{tab:participants} presents participant demographics, assigned platforms, device operating system, prior platform use, and self-rated familiarity with privacy and safety settings.

The study was approved by our Institutional Review Board, and before participation a parent or guardian provided consent while each participant provided assent through Adobe Sign. At the beginning of each session, the researcher reviewed the study procedure, voluntary participation, recording, confidentiality protections, and the participant's right to stop or skip a task.

\subsubsection{Think-Aloud Task Procedure}

The task sessions took place in November 2025, the same month as the expert evaluation, and lasted approximately 45 to 60 minutes. Each session opened with brief questions about prior platform use and familiarity with privacy and safety settings, which gave us context for interpreting task performance. Participants then completed a series of privacy and safety tasks on two platforms, with the assigned set drawn from the same six tasks and limited to the tasks available on the participant's platforms. Participants completed the tasks on research accounts that we provided, created with birth dates under 18 so that the accounts matched the age-based settings documented in the expert evaluation. 

Participants shared their phone screens through Zoom while working on their own devices, and the screen-shared interaction was recorded. We asked them to think aloud by describing what they were looking for and what they planned to select as they navigated. The researcher withheld assistance unless a participant requested a hint, became unable to proceed, or reached a technical or platform constraint that stopped the task. When assistance was given, we recorded whether it took the form of a general directional hint, the identification of a missed menu or control, or step-by-step prompting, and we asked participants to continue after a hint until they reached the endpoint or could go no further, so attempts completed with assistance include the interaction that followed the hint. 

\subsection{Analysis of Participant Task Performance}

\subsubsection{Unit of Analysis}
The unit of analysis was one participant's attempt to complete one task on one platform. We reconstructed each attempt from the Zoom screen recording and coded the interaction from the beginning to the end of the task. The sequence began when the participant first interacted with the phone to pursue the assigned task goal, and it ended when the participant reached the intended endpoint, stated that they were finished, clearly abandoned the attempt, the researcher ended or cancelled the task, or the session moved to the next task. Participants completed several tasks on the same platform, and each attempt began wherever the interface stood at the end of the previous task, so an attempt could begin from the app home or from a location within settings reached during an earlier task. Think-aloud narration included instances where the participant was actively attempting the task. 

\subsubsection{Interaction Cost Measures}

We coded the interaction cost of each task attempt as an interaction cost~\cite{Budiu2013-px, Lam2008-fh} using the click and interface state definitions from the expert evaluation, recording clicks, full screens, popups, layovers, and total interface states. We also considered the scrolls, completion status, and struggle level for every attempt. The codebooks for all of these measures is included in the appendix. Coding participant traces required rules the expert flows did not, so typing individual characters, swiping to scroll, and accidental touches without a visible response were not counted as clicks. Back navigation was counted as a click when it returned the participant to a different full screen, whether performed through an in-app control, a system button, or a back gesture. Interface states appearing without a corresponding tap, such as automatic transitions and system prompts, were counted as interface states but not as clicks, and a popup that expanded into a full page was counted as both states, with total interface states calculated as the sum of full screens, popups, and layovers. A scroll was one continuous swipe that visibly moved content vertically or horizontally, with repeated swipes counted separately; we counted scrolling regardless of whether it appeared goal-directed or hesitation-related, since both contributed to the interaction cost of completing the task. 

\subsubsection{Completion and Struggle Coding}
We coded a task as \emph{completed independently} when the participant reached the intended
endpoint without navigational guidance, as \emph{completed with assistance} when the participant
reached it after a directional hint, the identification of a missed control, or step-by-step
guidance, and as \emph{not completed} when the participant abandoned the attempt, ran out of task
time, or could not continue because the feature was unavailable in the tested context. Repeating
the task wording without naming a location or a control did not count as guidance. Struggle
records how the participant reached that outcome on four ordinal levels, where \emph{none} covers
a direct route with no observable hesitation, \emph{low} covers minor hesitation alongside smooth
independent progress such as a brief pause or a nearby menu check, \emph{moderate} covers
sustained searching with independent recovery through repeated scrolling, backtracking, wrong
turns, or rereading, and \emph{high} covers participants who became stuck, required navigational assistance, or never reached the endpoint. High struggle therefore overlaps with the assisted and
incomplete outcomes by construction, while struggle stays distinct from interaction cost, since a long but clear route produces many clicks with little difficulty and a short route can produce high struggle when the participant repeatedly selects the wrong option. 

\subsubsection{Coder Alignment and Visual Codebook}

Two researchers coded each sequence from three participants (P5, P6, and P7), covering 27 of the 88 coded attempts, and resolved every difference in counts and struggle ratings to consensus before the definitions were finalized. The alignment work gave particular attention to distinguishing full screens, popups, and layovers, determining where each attempt began and ended, separating scroll gestures from taps, identifying researcher assistance, and distinguishing moderate from high struggle. One researcher then selected exemplar clips of common interactions and edge cases and converted them into short GIFs illustrating clicks, scrolls, screen transitions, popups, layovers, and each struggle level, which the team used as a visual codebook to support consistent application of the written definitions. 
One researcher coded the remaining recordings using the finalized definitions, and the team discussed new edge cases before finalizing the affected attempts.

\subsubsection{Comparison with Expert Task Flows}
After coding all sequences for all participants, we compared each participant path against the expert path for the same task and platform. The comparison included two forms: a description of the paths used and any variance; and the interaction cost each path took. The first comparison appears in
each task section, which aligns participant navigation against the dark patterns coded on the expert flow, with wrong turns, near misses, looping, backtracking, deflected paths, researcher assistance, and incomplete attempts marked on the task flow figures. The second comparison appears in Section~\ref{sec:baseline-predict}, which indicates participant clicks and interface states beside the baseline figures for every task with a common endpoint. Notification management had no fixed endpoint, since participants decided which notifications to configure, so the difference between baseline and participant counts on that task registers how far a participant entered the configuration structure and does not measure efficiency along a shared route. Given the sample size, we report all counts descriptively and make no inferential claims about differences between platforms or
participants.

\section{Findings}
Sections~\ref{sec:screentime} through~\ref{sec:deletion} report four of the six tasks in
detail. We selected the tasks that produced the widest range of outcomes, covering the longest
documented routes (as in account deletion), and the largest differences between the expert
baseline and the cost participants incurred (as in notification settings). Table~\ref{tab:expert-baseline} reports the expert
baselines for all six tasks. Each section opens with the expert baseline that answers RQ1a. The
dark patterns coded along the same route follow, answering RQ1b, and the attempts participants
made on that task answer RQ2a. Section~\ref{sec:baseline-predict} answers RQ2b by placing the
participant attempts beside the expert baseline across tasks and platforms and identifying where
the two diverge. We provide a description of participants and tasks attempted on each platform in the appendix. 

\begin{table}[t]
\centering
\Description{Participant table, eleven rows by six columns, giving the participant identifier, age, age at first social media use, the two assigned platforms, device operating system, and self-rated familiarity with privacy and safety settings out of ten.}
\caption{Participant background, the two platforms used during the task session, and self-rated familiarity with privacy and safety (P\&S) settings on a 10-point scale. Participants ranged from 14 to 17 years old and completed tasks on different combinations of TikTok, Instagram, Snapchat, and YouTube. Self-rated familiarity is out of a scale of 10.}
\label{tab:participants}
\small
\begin{tabular}{llllll}
\toprule
\textbf{ID} & \textbf{Age} & \textbf{Age at First Use} & \textbf{Study Platforms} & \textbf{Device OS} & \textbf{P\&S Familiarity} \\
\midrule
P1  & 16 & about 13    & TikTok, Snapchat    & iOS     & 1--2 \\
P2  & 16 & 13--14      & YouTube, Instagram  & iOS     & 5 \\
P3  & 15 & 13          & Snapchat, Instagram & iOS     & 1 \\
P4  & 14 & 13          & TikTok, YouTube     & iOS     & 2--3 \\
P5  & 16 & 14          & TikTok, Instagram   & iOS     & 5 \\
P6  & 17 & ``few years ago''  & Snapchat, TikTok    & iOS     & 5 \\
P7  & 16 & 15          & Instagram, TikTok   & iOS     & 1 \\
P8  & 15 & 14--15      & TikTok, Instagram   & Android & 8 \\
P9  & 16 & 11          & Snapchat, TikTok    & iOS     & 7 \\
P10 & 16 & 12          & YouTube, Instagram  & iOS     & 7 \\
P11 & 15 & 13--14      & Instagram, YouTube  & Android & 3--4 \\
\bottomrule
\end{tabular}
\end{table}

\newcommand{\yy}{$\bullet$}                        
\newcommand{\nn}{$\circ$}                          
\newcommand{\na}{--}                               
\newcommand{\ph}{\textcolor{black!40}{\texttt{?}}} 

\newcommand{\taskblock}[2]{%
\begin{tabular}[t]{@{}c l cccc@{}}
\toprule
\textbf{Task} & \textbf{High-Level Dark Pattern} & \textbf{T} & \textbf{I} & \textbf{S} & \textbf{Y} \\
\midrule
\multirow{10}{*}{\rotatebox[origin=c]{90}{\tiny\shortstack{#1}}} #2
\bottomrule
\end{tabular}}

\begin{table*}[!ht]
\centering
\scriptsize
\setlength{\tabcolsep}{4pt}
\renewcommand{\arraystretch}{1.08}
\Description{Six blocks, one per task, each with one column per platform. Five rows per block record whether a high level dark pattern was coded along the path and five give the interaction cost. A filled circle means present, an open circle means absent, and a dash means the task is not available on that platform.}

\caption{Interaction cost required by each platform to complete the six privacy and safety tasks, reported as the minimum clicks and interface states along the documented path, together with the high-level dark patterns coded across that path using the Gray et al. ontology~\cite{gray2024ontology}. Interface states are reported as a total and by component type, distinguishing full screens, popups, and layovers, since flows of equal length differ in whether the underlying interface remains visible. TikTok (T), Instagram (I), Snapchat (S), and YouTube (Y).}
\label{tab:expert-baseline}

\begin{tabular}{@{}c@{}}

\taskblock{Task 1: Set a\\Screen-Time Limit}{%
 & Sneaking               & \yy & \yy & \na & \yy \\
 & Obstruction            & \yy & \yy & \na & \yy \\
 & Interface Interference & \yy & \yy & \na & \yy \\
 & Forced Action          & \nn & \nn & \na & \nn \\
 & Social Engineering     & \nn & \nn & \na & \nn \\
\cmidrule(l){2-6}
 & Number of Clicks       & 8 & 5 & \na & 4 \\
 & Interface States       & 7 & 5 & \na & 3 \\
 & \quad Full Screens     & 6 & 4 & \na & 1 \\
 & \quad Popups           & 0 & 1 & \na & 1 \\
 & \quad Layovers         & 1 & 0 & \na & 1 \\
}\hspace{6pt}%
\taskblock{Task 2: Set\\Private Account}{%
 & Sneaking               & \nn & \nn & \na & \na \\
 & Obstruction            & \yy & \yy & \na & \na \\
 & Interface Interference & \nn & \nn & \na & \na \\
 & Forced Action          & \nn & \nn & \na & \na \\
 & Social Engineering     & \nn & \nn & \na & \na \\
\cmidrule(l){2-6}
 & Number of Clicks       & 4 & 4 & \na & \na \\
 & Interface States       & 3 & 3 & \na & \na \\
 & \quad Full Screens     & 3 & 3 & \na & \na \\
 & \quad Popups           & 0 & 0 & \na & \na \\
 & \quad Layovers         & 0 & 0 & \na & \na \\
}\hspace{6pt}%
\taskblock{Task 3: Manage\\In-App Notifications}{%
 & Sneaking               & \nn & \nn & \nn & \nn \\
 & Obstruction            & \yy & \yy & \yy & \yy \\
 & Interface Interference & \yy & \yy & \yy & \yy \\
 & Forced Action          & \nn & \nn & \nn & \nn \\
 & Social Engineering     & \yy & \yy & \yy & \yy \\
\cmidrule(l){2-6}
 & Number of Clicks       & 40 & 83 & 27 & 16 \\
 & Interface States       &  6 & 22 &  6 &  3 \\
 & \quad Full Screens     & 5 & 21 & 6 & 3 \\
 & \quad Popups           & 0 & 0 & 0 & 0 \\
 & \quad Layovers         & 1 & 1 & 0 & 0 \\
}

\\[3ex]

\taskblock{Task 4:\\Report Content}{%
 & Sneaking               & \nn & \nn & \nn & \nn \\
 & Obstruction            & \yy & \yy & \yy & \yy \\
 & Interface Interference & \yy & \yy & \yy & \yy \\
 & Forced Action          & \nn & \nn & \nn & \nn \\
 & Social Engineering     & \nn & \yy & \nn & \nn \\
\cmidrule(l){2-6}
 & Number of Clicks       & 6 & 5 & 5 & 4 \\
 & Interface States       & 5 & 4 & 4 & 4 \\
 & \quad Full Screens     & 0 & 3 & 3 & 2 \\
 & \quad Popups           & 0 & 0 & 0 & 1 \\
 & \quad Layovers         & 5 & 1 & 1 & 1 \\
}\hspace{6pt}%
\taskblock{Task 5: Download\\Your Data}{%
 & Sneaking               & \nn & \nn & \nn & \nn \\
 & Obstruction            & \yy & \yy & \yy & \yy \\
 & Interface Interference & \nn & \nn & \nn & \nn \\
 & Forced Action          & \nn & \nn & \nn & \nn \\
 & Social Engineering     & \nn & \nn & \nn & \nn \\
\cmidrule(l){2-6}
 & Number of Clicks       & 7 & 9 & 5 & 4 \\
 & Interface States       & 7 & 9 & 5 & 4 \\
 & \quad Full Screens     & 5 & 7 & 5 & 0 \\
 & \quad Popups           & 1 & 0 & 0 & 0 \\
 & \quad Layovers         & 1 & 2 & 0 & 4 \\
}\hspace{6pt}%
\taskblock{Task 6:\\Delete Account}{%
 & Sneaking               & \nn & \yy & \nn & \na \\
 & Obstruction            & \yy & \yy & \yy & \na \\
 & Interface Interference & \yy & \yy & \yy & \na \\
 & Forced Action          & \yy & \yy & \yy & \na \\
 & Social Engineering     & \nn & \nn & \nn & \na \\
\cmidrule(l){2-6}
 & Number of Clicks       & 5 & 13 & 4 & \na \\
 & Interface States       & 5 & 11 & 3 & \na \\
 & \quad Full Screens     & 4 & 0 & 3 & \na \\
 & \quad Popups           & 0 & 0 & 0 & \na \\
 & \quad Layovers         & 1 & 11 & 0 & \na \\
}

\end{tabular}

\vspace{1.5ex}
{\scriptsize\textbf{Legend:} \yy~pattern present \quad \nn~pattern absent \quad \na~task not available on that platform  \quad T~--~TikTok \quad I~--~Instagram \quad S~--~Snapchat \quad Y~--~YouTube}

\end{table*}

\subsection{Setting a Screen Time Limit}
\label{sec:screentime}
\subsubsection{Expert Baseline}
Platforms frame screen time settings as a way for teens to remain in control of their attention and routines, yet the expert evaluation revealed substantial variation in how those settings were represented and configured across platforms. Establishing the same limit every day was straightforward, requiring between four and eight clicks across three to seven screens, whereas customizing the limit by day required additional clicks and screens, particularly on TikTok and YouTube. The added effort created a clear asymmetry between enabling a basic limit and tailoring it to different schedules such as weekdays and weekends. Further friction arose from differences in nomenclature, as Instagram placed related controls under \textit{Time Management}, YouTube used labels such as \textit{Time Watched} and break-related reminders, and TikTok relied on its own screen time and activity terminology. Consequently, conceptually similar controls were presented through different labels and menu structures, limiting the extent to which a route learned on one platform could be reused directly on another.

Moreover, multiple dark pattern strategies reinforced these control issues, as the screen time flows exemplified \textbf{Obstruction}\footnote{The notation uses \textbf{bold} for High-Level Pattern, \textit{italics} for Meso-Level Pattern, and \underline{\textit{underlined italics}} for Low-Level Pattern.} through \textit{Adding Steps}, with more detailed control requiring movement through multiple nested menus even under direct completion conditions. The absence of a consolidated grouping for related controls further aligned with \underline{\textit{Privacy Maze}}, requiring users to move across separate pages to locate and finalize parts of the same screen time task. In addition, the presentation of controls introduced \textbf{Interface Interference}, and on TikTok and YouTube, friendly illustrations and calm visual framing aligned with \textit{Emotional or Sensory Manipulation} by softening how actions intended to limit continued use were conveyed. TikTok's screen time flow also used \underline{\textit{False Hierarchy}}, making the route into screen time management visually clearer than the pathways for leaving or disengaging from the setting. Under the expert baseline, the control remained available, while more detailed or firmer configurations required greater interaction cost.

\subsubsection{Participant Performance}
\begin{figure*}[t]
  \centering
  \includegraphics[width=\textwidth]{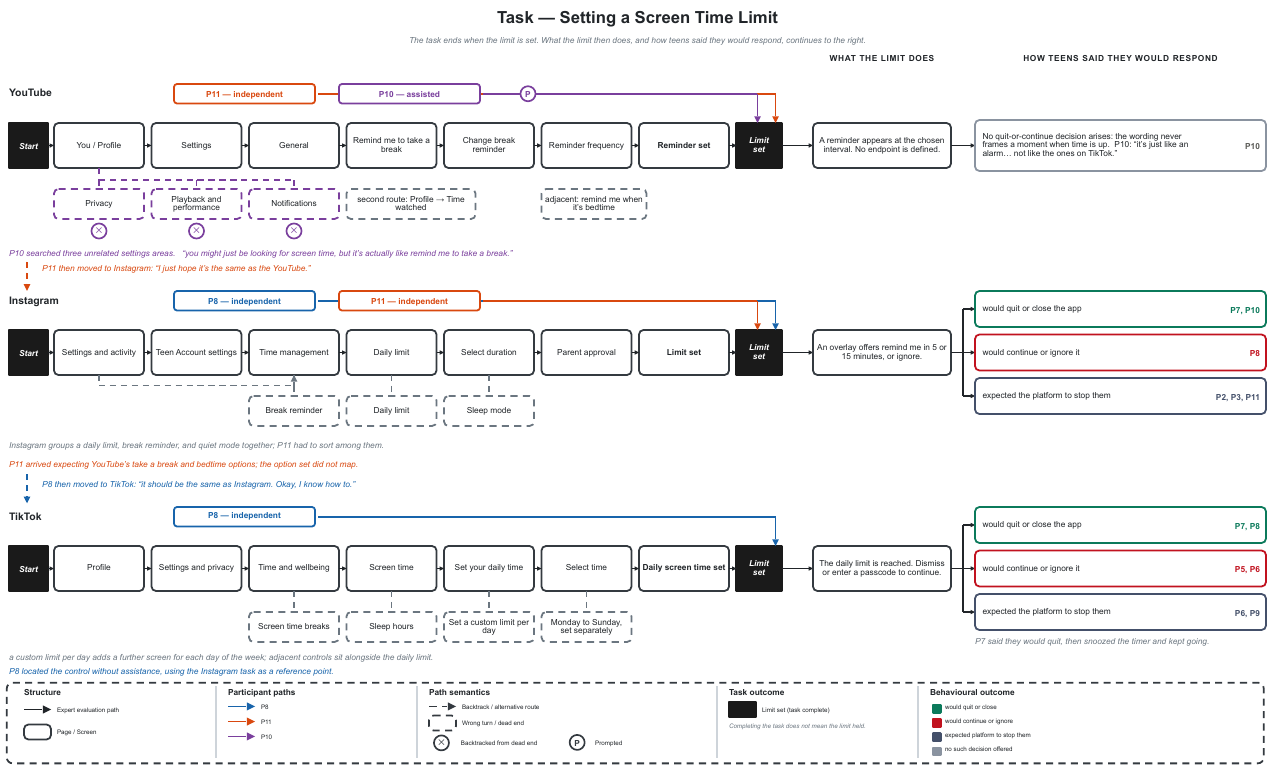}
  \Description{Three parallel task flow diagrams, one per platform, each showing the expert path with participant routes over it. The YouTube path runs Profile, Settings, General, Remind me to take a break, Change break reminder, Reminder frequency, Reminder set. The Instagram path runs Settings and activity, Teen Account settings, Time management, Daily limit, Select duration, Parent approval. The TikTok path runs Profile, Settings and privacy, Time and wellbeing, Screen time, Set your daily time, Select time. Dead ends and adjacent controls sit below each path. Participant routes and outcomes are described in Section~\ref{sec:screentime}, and per attempt counts are in Table~\ref{tab:detailed-performance}.}

  \caption{Setting a screen time limit across YouTube, Instagram, and TikTok.
  Each lane shows the expert path, with off-path screens below it and the dark
  patterns coded in the expert evaluation shown as spans, following the
  temporal analysis approach of Gray et al.~\cite{gray2025iliad}, and task flow analysis from Mildner et al. \cite{10.1145/3706598.3713776}. Dashed arrows
  between lanes mark where a participant carried a recently learned route to
  the next platform. The columns at right continue past task completion,
  showing what each platform does once the limit is reached and how
  participants said they would respond.}
  \Description{Three parallel task flow diagrams, one per platform, showing the expert path with participant routes over it. Each path starts at the platform home screen and ends when the limit is set. YouTube reaches it through Settings, General, and Remind me to take a break. Instagram reaches it through Settings and activity, Teen Account settings, Time management, and Daily limit. TikTok reaches it through Settings and privacy, Time and wellbeing, Screen time, and Set your daily time. Dead ends and adjacent controls sit below each path, and two columns at the right show what each platform does once the limit is reached.}

  \label{fig:screen-time-flow}
\end{figure*}
Participant attempts to set screen time limits varied both in how directly participants reached the control and in how they carried recently learned routes and terminology from one platform to another. Completing tasks on two platforms during the same session led several participants to approach the second platform by reusing what they had just learned on the first, making the cross-platform differences identified in the expert evaluation visible when similar controls appeared under different labels or settings structures. Figure~\ref{fig:screen-time-flow} reconstructs representative examples of the three resulting paths, which involved successful transfer, attempted reuse disrupted by different labels and options, and exploratory navigation through several plausible settings before reaching the target with assistance. Successful transfer occurred after P8 completed the screen time task on Instagram and began the same task on TikTok by saying, ``It should be the same as Instagram. I know how to. Let me try that.'' Using the Instagram task as a reference point allowed P8 to locate the corresponding TikTok control without assistance. P7 similarly remained close to the direct paths on both TikTok and Instagram, having already used screen time controls before the study. In these cases, differences between the platform structures did not substantially extend the participants' routes beyond the expert paths.

By contrast, attempted transfer became more difficult when platforms organized similar controls through different labels, options, or access points, as P11 completed the task on YouTube and began the Instagram attempt by saying, ``I just hope it's the same as the YouTube.'' The difference between the platforms became clearer when P11 described YouTube as providing primarily \textit{take a break} and \textit{bedtime} options, whereas Instagram presented a daily limit, break reminder, and quiet mode. Although P11 reached a similar type of control on both platforms, the route and available options mapped differently from one interface to the other, paralleling the platform-specific labels and menu structures identified in the expert evaluation. A more basic discontinuity in feature recognition emerged after P2 attempted the task on YouTube and responded to the Instagram task by asking, ``I thought that was only for YouTube. I can also do that on Instagram?'' Subsequent navigation through Instagram's settings led P2 to the control under \textit{Time Management}, even though experience with YouTube offered no recognizable label for locating the same broad function and reflected the nomenclature differences already present in the expert baseline. Terminology created a more sustained disruption during P10's YouTube attempt, as the search moved through settings areas including Privacy and prompted the following explanation of the mismatch between the task wording and the platform's label.
\begin{quote}
``You might get confused when I say screen time. You might just [be] looking for screen time, but it's actually like remind me to take a break.''
\end{quote}
P10 contrasted the YouTube terminology with TikTok's more explicit labeling, saying, ``On TikTok is boldly stated. But here just saying when you want to sleep.'' The YouTube attempt consequently extended across multiple settings areas and eventually required researcher guidance, with the expected screen time label failing to correspond directly to the platform's terminology. The difficulty extended beyond the depth of the settings hierarchy and reflected the interaction cost identified in the expert evaluation, where different names and locations made similar controls less recognizable across platforms. Moreover, P10 identified more than one route to related screen time controls within YouTube, returning to the profile area after locating \textit{Remind me to take a break} through Settings, pointing to \textit{Time Watched} as another route, and explaining that ``these two are actually like two ways of getting it.'' The multiple access points distributed one broad task goal across different labels and locations within the same platform, corresponding with the expert evaluation's finding that related controls were not always consolidated within one settings location. A related distinction appeared within TikTok when P6 encountered both the daily screen time limit and screen time breaks and asked, ``How different is that from the daily screen time?'' The adjacent controls served similar self-regulation goals, yet distinguishing between them still required additional navigation and interpretation and reflected the distributed organization identified in the expert baseline.

Exploratory navigation became more extensive when participants could not establish a workable route, as P9's TikTok attempt moved through \textit{Security and Permissions}, \textit{Family}, and \textit{Display} while searching for the screen time setting. After checking several plausible locations, P9 stopped and asked, ``I don't know. How do I do it?'' Researcher guidance was then required to locate the feature, while P10's YouTube attempt followed a similar pattern of searching through plausible settings before redirection toward the platform's break-related controls. By extending beyond the direct expert route and traversing multiple settings areas before reaching the intended control, both searches made the \underline{\textit{Privacy Maze}} identified in the expert evaluation visible in participant navigation. The additional searching also extended the \textbf{Obstruction} documented in the baseline, as participants accumulated further steps beyond those required by the direct platform path. Figure~\ref{fig:screen-time-flow} reconstructs representative examples of these paths, including a recently learned route that transferred across platforms, an attempted reuse that met different labels and options, and a search that moved through several plausible settings before reaching the target with assistance.

\subsection{Managing In-App Notifications}
\label{sec:notifications}
\subsubsection{Expert Baseline}
Platforms present notification settings as a mechanism for controlling when and how users are notified, yet the expert evaluation revealed substantial differences in the interaction cost of fully configuring those settings. Notification management was the most interaction-intensive task in the evaluation, requiring 40 clicks across 6 screens on TikTok, 83 clicks across 22 screens on Instagram, 27 clicks across 6 screens on Snapchat, and 16 clicks across 3 screens on YouTube.\footnote{We constrained the task to in-app notification management and did not consider any system-level notification controls.} The distribution of that interaction cost also differed across platforms, as TikTok and YouTube kept more changes within a simple notification settings space, allowing users to make multiple adjustments without repeatedly entering additional category screens, whereas Instagram and Snapchat distributed controls across multiple categories. Instagram and Snapchat fragmented settings across posts, messages, activity, and other system-generated alerts, requiring repeated clicks between category screens. 

The distributed structure of the notification flows reflected \textbf{Obstruction} through \textit{Adding Steps}, as reaching a fully customized state required repeated selections across individual settings rather than a consolidated control. Instagram's large number of granular settings further produced \textit{Choice Overload}. Many notification categories were also enabled by default, which aligned with \textbf{Interface Interference} through \textit{Bad Defaults}, because users seeking fewer alerts had to opt out repeatedly. Moreover, long lists of visually similar controls and the absence of a single overview contributed to \textit{Hidden Information}, particularly on Instagram, where the categories already reviewed or changed were difficult to identify at a glance. Beyond interface structure, socially meaningful categories and recommendation language around friends, messages, activity, and personalized content shaped how notification settings were presented. The expert evaluation coded such elements as forms of \textbf{Social Engineering} through \textit{Personalization}, with socially relevant cues framing the controls while the interaction cost of disabling notifications remained distributed across individual settings.

\subsubsection{Participant Performance}
Participant notification traces differed mainly in how far participants entered the configuration structure identified in the expert evaluation, while instructions to configure notifications according to individual preferences provided no single correct endpoint and made interaction cost conditional on the scope and persistence of each chosen configuration \cite{10.1145/3491102.3517493}. Trace length indicated configuration scope, with comprehensive configuration appearing only in routes that entered the broader category structure. At the shortest end, P8 reached Instagram's notification settings and left the existing configuration unchanged, saying, ``I usually don't touch it. I just leave it like that.'' Although the available categories remained accessible, P8's path ended near the top of the settings structure and retained the existing notification state, avoiding the repeated opt-out actions associated with \textbf{Interface Interference} through \textit{Bad Defaults} in the expert baseline. A second limited-scope route appeared when P10 used Instagram's temporary pause function during school and explained, ``If I'm in school, if I don't want my notification sound ringing and disturbing the class, I just pause it for maybe two hours, four hours.'' Selecting a temporary duration fulfilled P10's stated preference with little immediate interaction cost and bypassed the larger set of category-level decisions required for detailed configuration. The pause control offered a short branch through an architecture in which persistent category-level customization required substantially more interaction.

Selective persistent configuration moved farther into the available settings while remaining focused on categories relevant to each participant's stated goals, as P1 adjusted selected TikTok categories while explaining, ``Let's assume I don't want likes coming up. No comments. I want new followers, mentions, tags.'' P1's path covered several individual changes across a subset of the available categories. Meanwhile, P6 followed a similar approach on Snapchat by turning off friend suggestions, promotions, new posts, or live activity while leaving other notifications enabled. Across the selective paths, participants made fewer of the repeated selections that the expert baseline attributed to \textbf{Obstruction} through \textit{Adding Steps}. Further along the same continuum, P7 covered a larger portion of TikTok's available settings and explained, ``I usually want to decide on the particular notifications I get. So I want to make sure I don't get every notification.'' P7 then worked through multiple categories while deciding which notifications to keep or disable, and TikTok's concentration of controls within a primary notification space allowed several changes without repeated movement between separate category screens, keeping the route comparatively close to the centralized structure documented in the TikTok expert baseline.

Detailed and durable customization exposed progressively more of Instagram's fragmented notification architecture and associated interaction cost, beginning with P11's item-by-item review of the available options. P11 explained, ``If you want to get notifications that people liked your post, I think this is necessary. I'll activate them,'' before continuing through other categories, declining fundraiser notifications, and describing birthdays as useful ``for important people.'' The item-by-item process required repeated movement between category-level screens and exposed the \textit{Choice Overload} and \textit{Adding Steps} identified in the expert evaluation more directly than the shorter traces. Moreover, the socially differentiated options involving likes, birthdays, and fundraisers corresponded with the categories coded in the expert evaluation as \textbf{Social Engineering} through \textit{Personalization}. P10 followed a similarly detailed route on Instagram, continuing through multiple notification categories and making decisions about individual options as they appeared. P10's longer trace required repeated category entry, selection, and navigation back to the broader notification menu, revealing more of the configuration effort documented in the expert baseline. Continued movement between visually similar category screens, rather than configuration from a single consolidated view, also corresponded with the \textit{Hidden Information} identified in the expert evaluation.

\subsection{Reporting Content}
\label{sec:reporting}

\subsubsection{Expert Baseline}

Reporting content followed a similar multi-step sequence across platforms once the reporting control had been opened, requiring users to select the report option, choose a reason, and confirm the report. The principal platform difference appeared at the beginning of the flow, with TikTok and Snapchat using long-press interactions to reveal reporting controls while Instagram and YouTube placed them behind a three-dot overflow menu. Across both entry patterns, the sequential process reflected \textbf{Obstruction} through \textit{Adding Steps}, as completing the protective action required several screens and selections. Moreover, both entry patterns contributed to \textit{Hidden Information}, since reporting lacked a persistent primary control and users needed to know that a long press or overflow menu would expose it. Platform variation continued after submission, particularly on Instagram, where the interface presented an additional choice about whether the user still wanted to see the reported content after completing the reporting sequence. Completing the report on Instagram could therefore leave the reported content visible until the user made the additional viewing choice. The expert evaluation coded the additional decision as \textbf{Interface Interference} through \textit{Hidden Information}, since the relationship between submitting a report and removing or suppressing content remained unclear within the reporting control.

\subsubsection{Participant Performance}
\begin{figure}
    \centering
    \includegraphics[width=1\linewidth]{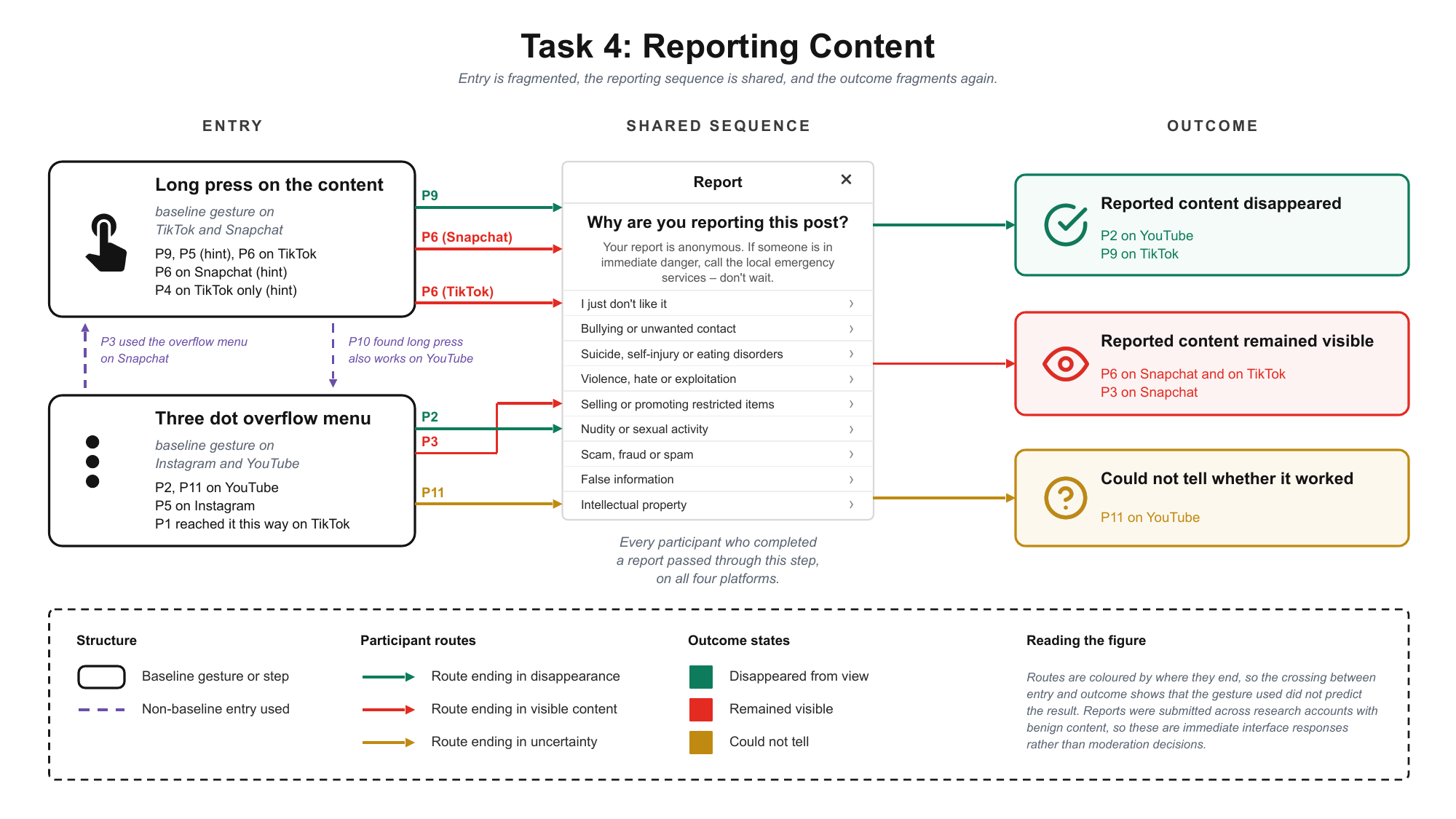}
   \Description{Three column diagram. The left column holds two entry gestures, a long press on the content and a three dot overflow menu. The middle column holds the reason selection panel that every completed report passed through. The right column holds three outcome states, the content disappearing, the content remaining visible, and the participant being unable to tell. Routes run left to right and cross between the outer columns, so the entry gesture did not predict the outcome.}

    \caption{Reporting content across the four platforms, drawn as three columns. The entry column
holds the two gestures that expose the report control, a long press on the content for TikTok
and Snapchat and a three dot overflow menu for Instagram and YouTube, and each participant
appears under the gesture they used. A dashed arrow marks an entry through the gesture that is
not the baseline for that platform, and the circled P marks where the researcher gave
navigational assistance. The middle column holds the reason selection sequence, which every
completed report passed through on all four platforms. The outcome column holds the state of
the reported content immediately after submission, which was disappearance from view, continued
visibility, or uncertainty about whether the report registered. Each route takes its colour from
where it ends, so a route that crosses between the entry column and the outcome column marks an
entry gesture and an outcome that do not correspond. Participants used research accounts and
reported benign content, so the outcome column records immediate interface responses and not
moderation decisions.}
    \label{fig:report-flow}
\end{figure}

Participant reporting attempts showed the greatest variation before participants entered the reporting flow, as difficulty centered on recognizing which interaction would reveal the report control. On TikTok, P4 searched for the report function without recognizing the long-press interaction and eventually stopped, saying, ``I don't know how to report it. I don't know how to report the video.'' P4 advanced only after the interviewer suggested long-pressing the video. Similarly, P5 said, ``report also? No, I can't find it,'' before reaching the reporting options through the content interaction menu. The P4 and P5 attempts surfaced the \textit{Hidden Information} that the expert baseline identified, since both participants searched for the report control without recognizing the gesture needed to expose it. By comparison, P9 used TikTok's long-press route independently but described the sequence as requiring several actions, saying, ``I had to like long press on the video and then like click it and then I had to go through all these steps.'' P9's independent entry still traversed the multi-stage sequence identified in the expert baseline as \textbf{Obstruction} through \textit{Adding Steps}. Meanwhile, P3 entered Snapchat's reporting flow through the three-dot menu instead of using the long-press interaction recorded in the expert baseline and explained, ``I clicked this three button,'' after the options appeared. On YouTube, P10 identified both available entry gestures and noted that ``you can also long press or use these three buttons by the side.'' The P3 and P10 routes differed from the single direct entry interaction recorded during the expert walkthrough while providing functionally equivalent access to the reporting options. P1 reached the same TikTok control without difficulty and described report as the final option among the initial actions available on the content, locating it at the end of a list assembled for other purposes. Across all entry routes, access remained dependent on a gesture or overflow interaction and lacked persistent display on the content itself, keeping the Hidden Information identified in the baseline relevant even when participants used different gestures.

Once participants exposed the report control, the reason-selection and confirmation screens were generally easier to follow, although the amount of information requested varied across platforms. Instagram presented long reason-selection menus that extended the reporting sequence beyond the initial content actions. Within the Instagram attempts, P2 paused at the reporting categories and remarked, ``There's a lot of options here now for me to like actually choose from," and P5 encountered the same extended reason-selection structure. YouTube required P4 to move through reason categories, more specific sub-options, a timestamp, and a free-text description before reaching the end of the sequence. The longer Instagram and YouTube routes extended the \textbf{Obstruction} through \textit{Adding Steps} identified in the expert baseline, with the number and type of required decisions varying by platform.

Post-submission responses created a second major source of variation across participants by producing continued visibility, disappearance, or uncertainty about whether the report had worked. On Snapchat, P6 completed the reporting flow and found the content still visible, saying, ``it just says thanks for reporting. It doesn't take it away. Unlike the block. The block takes it away.'' P6 further explained the expected outcome by saying, I think me reporting it should automatically take it away like the block one did.'' Meanwhile, P3 independently encountered the same visible outcome and remarked, ``Oh, so even if you report it, it's still here.'' Continued visibility after completion left P6 and P3 without an immediate visible change and made the uncertainty represented by \textbf{Interface Interference} through \textit{Hidden Information} in the expert evaluation visible in the participant attempts. By contrast, P2 observed the video disappear after reporting on YouTube, while P9 saw the reported content disappear on TikTok. P11 encountered a less clear YouTube outcome when the content remained visible after submission and said, ``I don't know if it's worked but I don't know if that's how it is.'' Compared with the platform-specific post-report state documented during the controlled expert walkthrough, the participant traces included disappearance, continued visibility, and uncertainty about whether submission had worked. The use of research accounts and generally benign content limits the observations to the immediate interface response after submission without establishing whether the platforms ultimately acted on the reports. Figure~\ref{fig:report-flow} reconstructs representative reporting traces that begin from different entry interactions, pass through the shared reason-selection flow, and end in whether the reported content remained visible immediately after submission.

\subsection{Deleting an Account}
\label{sec:deletion}

\subsubsection{Expert Baseline}

All platforms presented account deletion as an available account-management action, although the direct path to permanent deletion varied substantially. TikTok required a short path, whereas Snapchat exposed deletion through a more direct settings route that added an identity-confirmation step. Instagram required 13 clicks across 11 screens in the expert evaluation and introduced multiple decision points before permanent deletion. Across the three platform structures, the multi-page sequences reflected \textbf{Obstruction} through \underline{\textit{Adding Steps}}, as users moved through multiple screens before reaching the final deletion action. Moreover, TikTok, Instagram, and Snapchat required account verification before deletion could proceed. The expert flow documented verification as part of the deletion process, while the participant analysis distinguished credential entry from participant navigation failure.

Instagram's extended flow added several forms of \textbf{Interface Interference}, with the deletion control nested within Accounts Center and less permanent alternatives such as deactivation displayed prominently within the same sequence. Within the expert coding, the prominence of deactivation and the additional navigation required for permanent deletion aligned with \underline{\textit{False Hierarchy}} and \underline{\textit{Visual Prominence}}. Furthermore, the Instagram flow reflected \textbf{Sneaking} by presenting deactivation and other alternative actions after users entered with the intention to delete, while the repeated introduction of less final alternatives aligned with \underline{\textit{Bait and Switch}}. In addition, Instagram presented prompts involving the downloading or transferring of data and required users to select reasons for leaving, with ambiguous or strategically framed response options coded as \underline{\textit{Trick Questions}}.

Beyond the navigation sequence visible during the initial task, account deletion remained subject to a 30-day retention or waiting period across the evaluated platforms, delaying permanent deletion after the final request. The expert evaluation coded the delay as \textbf{Forced Action}, since submission initiated a waiting period and prevented immediate permanent deletion. Logging back into the account during the waiting period could cancel the deletion request and require the entire process to be restarted. The resulting reversibility extended the temporal length of the deletion flow beyond the screens visible during the initial task.

\begin{figure}
    \centering
    \includegraphics[width=1\linewidth]{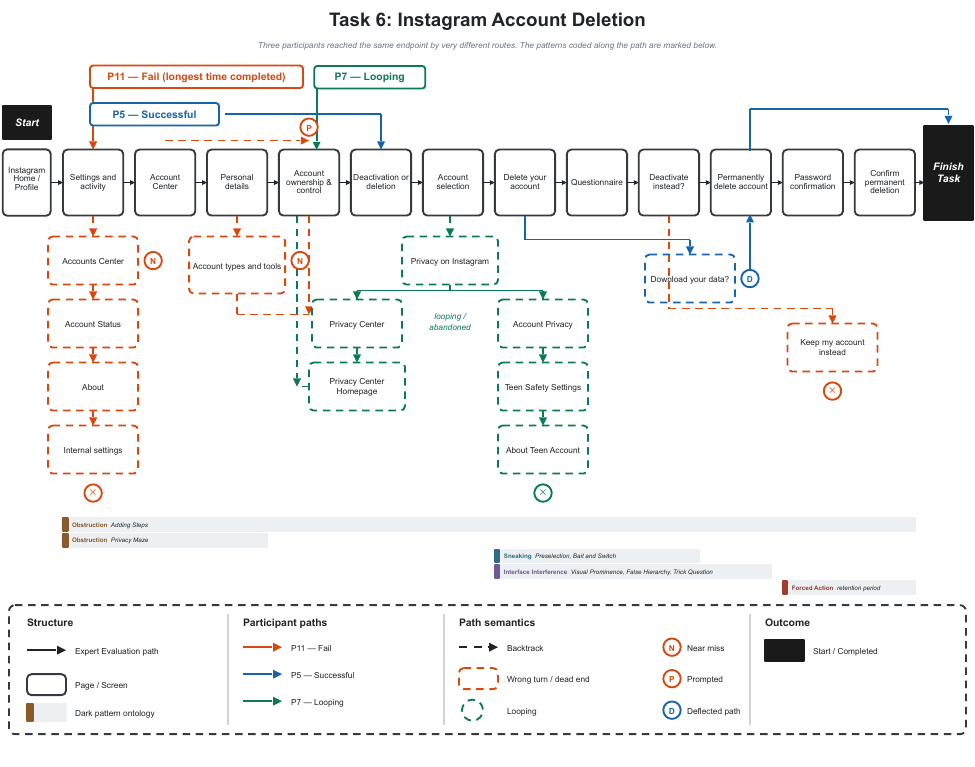}
    \Description{Task flow diagram of Instagram account deletion with three participant routes over the expert path. The path starts at the Instagram home screen and ends at confirmed permanent deletion, passing through Account Center, Personal details, Account ownership and control, Deactivation or deletion, a questionnaire, a Deactivate instead prompt, and password confirmation. Off path screens sit below. Five bars beneath show how far each coded pattern extends, with Obstruction across the early navigation, Sneaking and Interface Interference across the deactivation alternative, and Forced Action across the retention period.}

   \caption{Instagram account deletion, with three participant routes drawn over the expert
baseline. The row of boxes across the top is the baseline path from the Instagram home screen
to the completed deletion, and the boxes below it are screens that sit off that path. Each
route takes its own colour and the legend identifies it. The path semantics in the legend mark
what happened away from the baseline, covering backtracks, wrong turns that ended in dead ends,
loops through the same screens, near misses, points where the researcher gave navigational
assistance, and deflected paths where the platform offered an alternative action. The bars
beneath the diagram show how far each coded dark pattern extends, so a bar that spans several
boxes marks a pattern that persists across successive screens. The span representation follows
the temporal analysis of dark patterns in Gray et al.~\cite{gray2025iliad}, the pattern names
follow the ontology of Gray et al.~\cite{gray2024ontology}, and the mapping of a task flow from
entry point to endpoint follows Mildner et al.~\cite{10.1145/3706598.3713776}.}
    \label{fig:delete-flow}
\end{figure}

\subsubsection{Participant Performance}
Participant performance on account deletion varied most sharply on Instagram, where the expert evaluation identified the longest direct path and the greatest number of decision points before permanent deletion, and the three focal cases traced a progression from comparatively direct completion to extended searching and researcher-assisted navigation. The first case followed the route most closely aligned with the expert baseline, as P5 navigated through Instagram's account-management settings, reached \textit{Account Ownership and Control}, and identified \textit{Deactivate or delete} without a navigational hint. Even along the comparatively direct path, P5 moved through the alternatives and reason-selection screens documented in the expert evaluation. Reflecting on the task, P5 described the process as ``a little bit hard'' and attributed the difficulty to the fact that ``there's so many options'' and users must move through different reasons before reaching deletion. P5's attempt consequently remained close to the direct route while still encountering the \underline{\textit{Adding Steps}} present in the baseline flow.

The second case reached the deletion endpoint through substantially more movement across Instagram's settings structure, as P2 reasoned that deletion might be located in Accounts Center given its connection to account and personal information but moved back and forth through the interface before finding the relevant control. P2's attempt recorded the highest click count among the Instagram deletion traces in the dataset. After entering the deletion sequence, P2 paused at the distinction between deactivation and permanent deletion and initially considered deactivation before continuing. P2's hesitation occurred at the same decision point where the expert evaluation identified \underline{\textit{False Hierarchy}} and \underline{\textit{Visual Prominence}}, with the less permanent alternative presented prominently alongside deletion.

The third case showed the greatest divergence from the expert path in P7's attempt, while P8 and P11 provided comparable evidence of difficulty locating the deletion entry point. Although P7 had previously deleted an Instagram account during personal use, P7 could not reconstruct the route during the study and searched through settings and privacy-related areas before saying, ``I actually can't find it.'' When asked about the earlier experience, P7 responded, ``No, I don't remember.'' P7 further connected the search to the deletion task completed on TikTok by explaining, ``The account deletion process is what I did on TikTok that I wanted to do on Instagram.'' The route that had worked on TikTok did not map directly onto Instagram's Accounts Center structure, and P7 eventually needed a hint directing the search toward Accounts Center and Personal Details before continuing into the deletion sequence. Repeated entry into plausible privacy and account settings extended the \textbf{Obstruction} identified in the expert baseline through additional searching and backtracking. Similarly, P8 searched through settings, commented that ``it's a bit difficult,'' and later said, ``I think I made a mistake'' after entering an unrelated area. P8 could not locate the deletion path independently and required step-by-step guidance. Meanwhile, P11 searched through settings and privacy-related pages before receiving assistance and later described ``deleting the account on Instagram'' as the task where P11 ``quit'' and ``had a lot of difficulties going through that.'' After reaching the deactivation and deletion screen, P11 stopped to compare the two options before proceeding. P11's path therefore involved difficulty first in locating the deletion entry point and then in evaluating the alternatives presented within the deletion sequence. The later hesitation occurred at the same decision point where the expert evaluation identified \underline{\textit{False Hierarchy}} and \underline{\textit{Visual Prominence}} around the prominence of deactivation relative to permanent deletion.

Across the Instagram cases, alternative actions embedded in the expert flow remained visible during otherwise successful attempts. Participants encountered prompts involving deactivation, data download, and reasons for leaving before reaching permanent deletion, and P5 encountered the additional options despite locating the deletion route independently. The added screens lengthened the path after entry into the deletion sequence and corresponded with the \textbf{Sneaking} and \underline{\textit{Bait and Switch}} identified in the expert evaluation, even when participants continued toward completion. Evidence for \underline{\textit{Trick Questions}} required consistent participant misunderstanding or redirection caused by the wording of the responses, and the reason-selection traces alone did not meet that threshold.
Compared with the pronounced variation on Instagram, TikTok and Snapchat attempts generally remained closer to their shorter expert baselines. On TikTok, P4 moved from the profile to the account menu and immediately recognized \textit{Deactivate or delete account}, while P9 similarly located the control and proceeded toward permanent deletion. P5 found the TikTok deletion option while scanning the account-information menu, and P7 later described the TikTok process as ``straightforward.'' The TikTok paths involved fewer exploratory detours before participants entered the deletion sequence. Snapchat attempts were similarly compact, with P1 entering Settings, scrolling until locating \textit{Delete Account}, and proceeding through the task. P9 initially said, ``I can't find it,'' but located the option shortly afterward and continued without an extended search across unrelated settings. The additional identity-confirmation step documented in the expert baseline remained part of the Snapchat flow, while wayfinding remained more compact than the navigation observed around Instagram's Accounts Center. Figure~\ref{fig:delete-flow} compares the three Instagram traces to the expert path, showing a comparatively direct completion, an extended route that reached the same endpoint after several screens that led nowhere, and a route that required researcher guidance before the participant could continue into the deletion sequence.

\subsection{What the Expert Baseline Does and Does Not Predict}
\label{sec:baseline-predict}

Table~\ref{tab:teen-vs-expert} places the participant attempts beside the expert baseline for the same
task and platform. The expert baseline records how far a control sits from the platform home
screen for an evaluator who already knows its name and its entry point, but does not consider how recognizable the route is. The three comparisons below separate the tasks
where distance governs participant interaction cost from the tasks where recognition governs it.

\subsubsection{Participant Interaction Is Often Higher, But Not Always}

Participant interaction cost rises above the expert baseline on the tasks where participants had to
locate an unfamiliar control, and falls below it on the task where participants stopped short of
the configuration the platform requires for meaningful protection. Screen time and deletion sit above the baseline on every
platform, with TikTok deletion at 10.6 clicks compared to 5, Instagram deletion at 19.7 compared to 13,
and YouTube screen time at 13.0 compared to 4. 
Notification configuration sits below the
baseline on all four platforms, and on Instagram participants recorded 4.6 clicks on average
against the 83 the platform requires to fully disable notifications. The participant notification attempts also
differ from the expert route in composition, since participants recorded at most 2.1 clicks for
each interface state while the expert route concentrates repeated selections inside a small
number of screens, which marks movement through the settings structure instead of configuration
inside a category screen. Reporting on Instagram and deletion on Snapchat are both within one click
of their baselines. 

\subsubsection{Range Is Shaped By Recognition}

The range distinguishes an interaction cost that every participant incurred from an interaction
cost that depended on whether a participant recognized the route. Deleting an account on TikTok
averaged 10.6 clicks against a baseline of 5, and every attempt fell between 9 and 12. The narrow
spread identifies a route longer than the expert path that all participants walked, with route
finding contributing little variation between them. Reporting content on Snapchat averaged 11.3
clicks across a range of 5 to 24. Some participants used the long press documented in the expert
path and finished near the baseline, while others searched before the reporting flow opened, so
one average covers two different experiences of the same interface. Setting a screen time limit
on YouTube and deleting an account on Instagram show comparable spread. A narrow range identifies
an interaction cost the platform imposes on every participant, and a wide range identifies
variation in how legible the route was.



\begin{table*}[t]
\centering
\Description{Comparison table grouped by task and then platform. Two columns give the expert baseline as clicks and interface states, four give the participant average, standard deviation, range, and average interface states, and a final column gives the number of attempts.}

\caption{Interaction cost by task and platform, comparing the expert baseline with participant attempts. \textit{Clicks} and \textit{Screens} report the clicks and interface states along the expert path from the platform home screen, established through the expert evaluation when the goal, entry point, and route are already known. Participant attempts began from the interface state left by the preceding task, so an attempt can fall below the expert figure without having taken a shorter route to the control. \textit{Avg Clicks}, \textit{SD}, \textit{Min--Max}, and \textit{Avg Screens} report the mean, standard deviation, range, and mean number of interface states across the participants who attempted that task on that platform, and \textit{N} reports how many participants attempted it. Interface states are the sum of full screens, popups, and layovers. The two halves are not estimates of the same quantity, since the expert baseline is a minimum along one path while the participant columns are means across attempts, so the difference between them is the object of analysis rather than an error term. Cells rest on between three and seven attempts, and the range indicates whether interaction cost above the baseline was distributed across participants or concentrated in a few extended searches. Tasks not assigned on a platform are omitted from the table.}
\label{tab:teen-vs-expert}
\begin{tabular}{@{}l l r r r r r r r@{}}
\toprule
\multirow{2}{*}{\textbf{Task}} & \multirow{2}{*}{\textbf{Platform}}
  & \multicolumn{2}{c}{\textbf{Expert Baseline}}
  & \multicolumn{4}{c}{\textbf{Participant Attempts}}
  & \multirow{2}{*}{\textbf{N}} \\
\cmidrule(lr){3-4} \cmidrule(lr){5-8}
 & & \textbf{Clicks} & \textbf{Screens} & \textbf{Avg Clicks} & \textbf{SD} & \textbf{Min--Max} & \textbf{Avg Screens} & \\
\midrule
\multirow{3}{*}{Set Screen Time Limit}& TikTok    &  8 &  7 & 12.1 & 5.6 &  4--22 & 10.7 & 7 \\
 & Instagram &  5 &  5 &  8.0 & 3.5 &  4--14 &  5.5 & 6 \\
 & YouTube   &  4 &  3 & 13.0 & 7.2 &  6--21 & 15.3 & 4 \\
\midrule
\multirow{4}{*}{Manage In-App Notifications}
 & TikTok    & 40 &  6 &  6.2 & 3.9 &  1--10 &  3.0 & 6 \\
 & Instagram & 83 & 22 &  4.6 & 2.3 &   2--7 &  4.2 & 5 \\
 & Snapchat  & 27 &  6 &  5.7 & 1.5 &   4--7 &  4.7 & 3 \\
 & YouTube   & 16 &  3 &  4.0 & 2.6 &   2--7 &  2.0 & 3 \\
\midrule
\multirow{4}{*}{Report Content}
 & TikTok    &  6 &  5 &  7.2 & 2.8 &  4--11 &  6.5 & 6 \\
 & Instagram &  5 &  4 &  6.1 & 2.7 &  3--11 &  6.4 & 7 \\
 & Snapchat  &  5 &  4 & 11.3 & 11.0 & 5--24 &  7.0 & 3 \\
 & YouTube   &  4 &  4 &  9.5 & 5.3 &  4--15 &  8.5 & 4 \\
\midrule
\multirow{3}{*}{Delete Account}
 & TikTok    &  5 &  5 & 10.6 & 1.3 &  9--12 &  9.9 & 7 \\
 & Instagram & 13 & 11 & 19.7 & 7.6 & 14--34 & 16.5 & 6 \\
 & Snapchat  &  4 &  3 &  4.8 & 1.0 &   4--6 &  5.3 & 4 \\
\bottomrule
\end{tabular}
\end{table*}

\section{Discussion}

As we have demonstrated, the presence of a privacy or safety setting does not make it usable or durable. In this section, we first consider how social media platforms might consider creating a protected \textit{experience} instead of a set of features. We then propose how our wayfinding audit might be constructively extended in future work to proactively identify flaws and support regulation of platforms. 

\subsection{Moving from Feature Availability to a Protected Experience}
\label{sec:beyond-availability}

Platforms present privacy, safety, and wellbeing settings as mechanisms that protect the people who use them. Formal availability, however, does not establish that a setting provides protection in practice. 
When protection depends on user action, a setting must be usable: teenagers must be able to 
recognize that a control exists, locate it, understand the action it offers, configure it in a protective way, and interpret the outcome. Protective defaults can reduce the need for this work, but they must still make the resulting state understandable and verifiable. Whether activated by default or configured by the user, the safeguard must also be durable: its protection should persist long enough to be meaningful rather than being easily overridden or undermined by less-protective defaults.

Our findings demonstrate that the settings teenagers are expected to rely upon fail on basic usability grounds due to the presence of dark patterns, a lack of protective default measures, and a lack of understanding of what these settings can offer the user. Mere availability of a setting, often as an opt-in rather than opt-out, does not provide usable protection when a teenager cannot locate a setting, understand the available action, or complete the intended configuration. Our participants encountered unfamiliar terminology, hidden entry points, and confusing endpoints---thus hampering their ability to be properly protected on the platform. Search and interpretation substantially extended some tasks well beyond the documented route, particularly when participants had to recognize an unfamiliar label or entry gesture before they could begin the required sequence. 
Options that preserved engagement also frequently remained easier to reach than the more protective option. 
However, an apparently short interaction did not necessarily indicate efficient completion even though its interaction cost may appear lower. In the notification tasks, attempts sometimes remained below the expert baseline because participants retained an existing default, selected a temporary alternative, or stopped before completing the configuration the task required. Evaluating usability therefore requires attention to the endpoint and resulting state, not only the number of actions recorded.

Importantly, these usability problems observed across the tasks reproduce failures long known and addressed by foundational HCI principles. Unfamiliar labels weaken the match between a system and the language teenagers use for a protective goal, hidden gestures demand recall where the interface could support recognition, and inconsistent navigation across services undercuts consistency and standards~\cite{nielsenmolich1990, nielsen1994heuristics}. Endpoints that provide little or no confirmation weaken visibility of system status because users cannot tell whether the action took effect. The same protective goal also translates into different actions on each service, which increases the labor needed for the user to achieve their goal Without a basic adherence to usability standards, it is perhaps questionable whether the settings were meant to perform their claimed function at all---as underscored in recent testimony from Meta witnesses as part of US State action against the company that showed a lack of protective defaults and very low adoption of key features such as screen time controls ~\cite{aljazeera2026mosseri}.

While usability is important and achievable, HCI knowledge can also be leveraged to set a new bar for protection on social media platforms. Basic HCI principles direct evaluation toward the design and likely use of a safety feature, but additional attention to the lived experience of teens~\cite{Gairola2026-do,Sanchez-Chamorro2024-hv} could also encourage designers and platforms to consider how to ensure a truly \textit{protective experience} for teens and other vulnerable groups. To realize this goal, assessments should consider not only whether a required setting is technically present, but whether it functions as a meaningful safeguard for the people expected to rely on it. The following section further describes the wayfinding audit pioneered in this study as a foundation for considering both usability and the lived experience of users as they engage with privacy, safety, and wellbeing settings.

\subsection{Beyond Usability: A Preliminary Wayfinding Audit}
\label{sec:wayfinding-audit}

Evaluating a protected experience requires a technique that follows a protective goal from the point a user forms it to the point the interface confirms an outcome. We describe the procedure as a preliminary wayfinding audit, an auditing technique that combines established forms of basic usability testing and expert evaluation. 
Participant sessions relied upon moderated usability testing in which our participants pursued defined protective goals, while the expert evaluation drew on established practices such as the cognitive walkthrough and dark patterns analysis by documenting the actions and interface states associated with the same goals~\cite{lewis1990walkthrough, wharton1994cognitive, gray2025iliad}. The expert path serves as an idealized comparison case because the evaluator begins with the goal, entry point, and sequence already known, so it records what the platform affords and carries no claim about optimal design or about how competently our participants performed. The two paths and their endpoints give the audit three points of reference for a single task, covering what the platform requires, what a user encounters, and what state an attempt actually reached.

Each component answers a narrower question than our argument requires. Basic usability testing establishes whether our participants could complete a protective task and where an attempt broke down, providing no insights about which parts of the interface serve the platform's interest in continued engagement. Lam defines interaction cost through moments when dialogue between a user and a system breaks down or the interface becomes an obstacle that the user must overcome~\cite{Lam2008-fh}, which makes the expert path and the participant path comparable while leaving any difference between them unexplained. Dark patterns analysis~\cite{gray2025iliad} supplies an explanation by locating where the platform favors an engaging outcome over a protective one, but it does not directly measure the effort that design imposes. 
Using this range of methods allows us to build a fuller picture of the functionality from multiple perspectives. A participant path shorter than the expert path would count as efficient if interaction cost were the only measure, and the endpoint shows instead that our participants retained a default, selected a temporary alternative, or stopped before completing the configuration the task required. Added work also stays legitimate when it protects an account or supports an informed decision, and the dark patterns analysis separates work of that kind from work that steers our participants toward the outcome the platform prefers.

The methodological contribution lies in adapting these familiar HCI methods to a current auditing need that extends beyond whether a safety feature \textit{exists} to whether users can \textit{find it, use it, and interpret the outcome}. Task flow diagrams, usability sessions organized around defined tasks, and comparison of user activity with the expert path are established parts of the procedure, none of which are novel in their own right. The diagrams also adopt the Temporal Analysis of Dark Patterns (TADP) approach in Gray et al.~\cite{gray2025iliad}, allowing identification of instances where obstruction or another pattern persists across successive screens, which lets the audit report where along a route a pattern appears and how long it impedes a user's goal. Assembling these methods into a single methodology also surfaced the conditions that make a protective state hard to sustain, including temporary alternatives, defaults that preserve engagement, and endpoints that give a user no way to confirm what changed.

The audit lays the groundwork for flexible evaluation of privacy and safety settings. 
The present study demonstrates the procedure within the examined platforms and tasks, and other contexts may need a different balance among usability testing, expert evaluation, and dark patterns analysis. Possible contexts include withdrawing consent, ending a subscription, changing a data sharing preference, reporting harm, and leaving an account, where the goals, risks, and confirmation requirements differ enough to change what makes a useful comparison. Designers can run the audit before deployment to find where a protective feature stays hidden, is more difficult to use than its purpose warrants, or becomes less durable. Task flow diagrams may also make navigation, branching, and the duration of coded patterns inspectable to auditors and regulators, who need evidence about how a mandated control behaves for the people expected to use it. Future development includes testing the procedure across platforms and design versions, specifying comparisons when services organize the same goal differently, establishing which components are necessary for which protective tasks, and working out how the audit's task flow representation aligns with TADP as that framework develops.

\section{Implications and Future Work}
\label{sec:implications-future}

The measures and the comparison are likely to transfer beyond the population and the platforms we studied. Clicks,
interface states, and scrolling actions carry no assumption about who performs them, and a reconstructed
route set against an observed attempt holds wherever a protective action sits behind nested
navigation. Older adults ~\cite{Sanchez_Chamorro2024-nn}, people who manage an account for someone else, and people who joined a
platform recently bring different prior knowledge of where a route leads, so the distance between
the required cost and the incurred cost would register the difference directly. The same
evaluation approach could be used on operating systems, games, and other applications teenagers use, where a
protective control can be equally available and equally hard to find, and we encourage usability
work that measures a control against what teenagers expect to find and where they expect to find
it.

We only engaged with each participant during one sessions and these usability tests were not designed as interventions, yet they frequently functioned as
incidental privacy education. 
Participants accessed features they had never seen, built interpretations for
settings they had ignored, and named specific features they intended to adopt afterward. P9, on discovering the ``download my data'' function, said ``I didn't even know this was a thing, but I feel like it's a good function'' and gave an accurate account of its purpose within a minute, while P11 reported learning ``some certain things'' and P2 described the session as ``educative.'' The barrier appears to be neither disinterest nor incapacity, since the features are not encountered during ordinary platform use. Future work could examine how teenagers could engage with privacy and safety controls before a moment of need arrives, and whether surfacing a control where it is relevant to what a teenager is already doing removes the searching our traces identify as the costly part of the task. Researchers should consider how situated encounters might be placed within platforms, schools, or family guidance tools without displacing teen agency. 

\section{Conclusion}
\label{sec:conclusion}

We paired an expert reconstruction of six privacy and safety task flows across TikTok, Instagram, Snapchat, and YouTube with think-aloud sessions in which 11 teenagers attempted the same tasks. Most attempts cost more than the documented route requires, and the excess accumulated before participants entered the route, while notification attempts cost far less because participants stopped short of the configuration the platform structure demands. Dark pattern strategies recurred on every platform we examined while the routes to the protective controls shared nothing but a goal, so knowledge of what raises interaction cost carries across services while knowledge of how to reach a control does not. Evaluating a protective control therefore requires asking whether its route can be found and whether the resulting state can be verified. We offer the wayfinding audit as an initial articulation and demonstration of a method for evaluating a protected experience. Its contribution comes from the integration of established HCI methods into one procedure that connects interface design, dark patterns, interaction cost, observed use, and the durability of a protective outcome.

\bibliographystyle{ACM-Reference-Format}
\bibliography{sample-base}

\appendix

\section*{Appendix: Coding Scheme for Task Interaction and Completion}
\setcounter{table}{0}
\renewcommand{\thetable}{A\arabic{table}}

\subsection*{A. Interaction Metrics}
\scriptsize
\Description{Definition table for the seven interaction measures, giving each element's operational definition, when to count it, when not to count it, and the edge case rules.}
\begin{longtable}{@{}p{1.5cm} p{3.0cm} p{3.0cm} p{2.6cm} p{2.5cm}@{}}
\caption{Interaction metric definitions used to code teen task attempts.}\\
\label{tab:interaction-metrics} \\
\toprule
\textbf{Element} & \textbf{Operational Definition} & \textbf{Count It When} & \textbf{Do Not Count It When} & \textbf{Edge Case Rules} \\
\midrule
\endfirsthead
\multicolumn{5}{c}{\tablename\ \thetable\ -- \textit{Continued from previous page}} \\[4pt]
\toprule
\textbf{Element} & \textbf{Operational Definition} & \textbf{Count It When} & \textbf{Do Not Count It When} & \textbf{Edge Case Rules} \\
\midrule
\endhead
\midrule
\multicolumn{5}{r}{\textit{Continued on next page}} \\
\endfoot
\bottomrule
\endlastfoot

Time Used & Elapsed time spent actively attempting the assigned task, including think aloud speech that occurs while the participant continues interacting with the interface & Start when the participant first interacts with the phone to pursue the assigned goal. End when the participant completes the task, clearly abandons the attempt, states that they are finished, or the researcher ends the task & Do not count post-task discussion, debriefing, or explanation after the task attempt has ended & Use this measure only for attempts for which the complete task interval can be reconstructed from the recording \\[4pt]

Full Screen & A full-page transition to a new and meaningfully distinct interface state & The participant opens a new page, enters a settings sub-page, or uses back navigation to return to a different full-page state & The participant remains on the same page and only scrolls, or a nonblocking toast or banner appears & Back navigation counts when it produces a different full-page state. Do not count researcher-requested revisiting after the task has ended \\[4pt]

Popup & A centered modal dialog displayed over the current interface that requires a response before the participant can proceed & The participant must confirm, choose, or dismiss an alert or modal dialog & A nonblocking toast or banner, a bottom sheet or layover, or a full-page transition & If a popup expands into a full-page view, count one popup followed by one full-screen transition \\[4pt]

Layover & A bottom sheet, popover, or overlay panel displayed over the current screen while the underlying interface remains visible & A panel slides over or rises from the current page and requires interaction to continue or close & A small toast or banner, a centered modal dialog, or a full-page replacement & Retain layovers as a separate state type. Do not collapse them into popups when reporting participant-level traces \\[4pt]

Total Interface States & The total number of distinct interface states traversed during the task attempt & Sum all coded full screens, popups, and layovers for the attempt & Do not add scrolls, clicks, or nonblocking transient messages & Report the component counts as S, P, and L where space permits; total interface states equal S + P + L \\[4pt]

Click & A discrete tap that triggers an interface response & The participant taps to open a menu or page, toggle a setting, select an option, navigate back, confirm a choice, or dismiss a modal & Swipes used for scrolling, typing characters, or accidental touches that produce no visible response & A tap that opens a popup or layover counts as one click. A later selection within it counts as another click \\[4pt]

Scroll & One continuous swipe gesture that visibly moves content vertically or horizontally & Any completed swipe moves the page, list, carousel, or panel, including short flicks & A touch that does not move content or a tap used to select an item & Each continuous swipe counts as one scroll. Repeated swipes count separately. Hesitation is coded through the struggle rating and, where available, through notes on repeated scrolling, reversal, rereading, or pauses \\

\end{longtable}
\normalsize

\subsection*{B. Completion}
\scriptsize
\Description{Definition table for the three completion categories, giving each one's definition, when to assign it, and what should not be treated as assistance or failure.}
\begin{longtable}{@{}p{2.0cm} p{3.4cm} p{3.6cm} p{3.4cm}@{}}

\caption{Completion coding definitions used to assess teen task performance.}\\
\label{tab:completion-coding} \\
\toprule
\textbf{Completion} & \textbf{Definition} & \textbf{Count It When} & \textbf{Do Not Treat as Assistance or Failure} \\
\midrule
\endfirsthead
\multicolumn{4}{c}{\tablename\ \thetable\ -- \textit{Continued from previous page}} \\[4pt]
\toprule
\textbf{Completion} & \textbf{Definition} & \textbf{Count It When} & \textbf{Do Not Treat as Assistance or Failure} \\
\midrule
\endhead
\midrule
\multicolumn{4}{r}{\textit{Continued on next page}} \\
\endfoot
\bottomrule
\endlastfoot

Completed independently & The participant reaches the intended task endpoint without navigational guidance from the researcher & The participant finds the relevant feature, completes the required action, and does not receive a hint about where to go or what to select & Repeating the task wording, supplying a required credential or verification code, or resolving a technical issue outside the participant's control \\[4pt]

Completed with assistance & The participant reaches the intended task endpoint after receiving one or more navigational hints or prompts & The researcher redirects the participant, identifies a menu or control, points out a missed option, or provides step-by-step help, after which the participant finishes & The final action may still be performed by the participant, but the attempt is not counted as independent completion \\[4pt]

Not completed & The participant does not reach the intended task endpoint before the attempt ends & The participant abandons the task, the researcher ends or cancels the task, the feature is unavailable in the tested context, or the recording does not establish successful completion & Do not use this category solely because a credential or verification step could not be completed; document such cases separately as platform or study constraints \\

\end{longtable}
\normalsize

\subsection*{C. Struggle Level}
\scriptsize
\Description{Definition table for the four struggle levels, giving each one's operational definition, observable indicators, and the completion outcomes it corresponds to.}
\begin{longtable}{@{}p{1.5cm} p{3.6cm} p{4.5cm} p{2.8cm}@{}}
\caption{Struggle level definitions used to characterise the degree of difficulty participants experienced during each task.}\\
\label{tab:struggle-levels} \\
\toprule
\textbf{Level} & \textbf{Operational Definition} & \textbf{Observable Indicators} & \textbf{Completion Relationship} \\
\midrule
\endfirsthead
\multicolumn{4}{c}{\tablename\ \thetable\ -- \textit{Continued from previous page}} \\[4pt]
\toprule
\textbf{Level} & \textbf{Operational Definition} & \textbf{Observable Indicators} & \textbf{Completion Relationship} \\
\midrule
\endhead
\midrule
\multicolumn{4}{r}{\textit{Continued on next page}} \\
\endfoot
\bottomrule
\endlastfoot

None & The participant follows the expert baseline or a functionally equivalent direct path with no observable hesitation & Baseline or near-baseline clicks and interface states, little or no scrolling, no wrong turns or backtracking, and no visible pause to decide & Completed independently \\[4pt]

Low & The participant shows minor hesitation but progresses smoothly and independently & A brief pause, one nearby menu check, one or two extra scrolls, or a small number of extra clicks or interface states, followed by quick recovery & Completed independently \\[4pt]

Moderate & The participant shows sustained searching or looping but independently recovers and reaches the endpoint & Repeated scrolling, backtracking, one or more wrong turns, repeated rereading, or clearly elevated clicks and interface states, without a navigational hint & Completed independently \\[4pt]

High & The participant becomes stuck, requires navigational assistance, abandons the attempt, or cannot reach the endpoint & An explicit request for help, researcher redirection or step-by-step prompting, prolonged unsuccessful search, task cancellation, or failure to reach the endpoint & Completed with assistance or not completed \\

\end{longtable}
\normalsize

\subsection*{D. Detailed Task Performance by Participant, Platform, and Task}
We distinguish among full-screen transitions (S), popups (P), and layovers (L); total interface states equal S + P + L. A state type is reported only where the participant encountered it, so a trace recorded as popups alone indicates that no full-screen transition occurred during the attempt. Completion uses three categories: completed independently, completed with assistance, and not completed. High struggle is assigned whenever navigational assistance was required or the participant did not reach the endpoint. The Basis for Rating column records the observable grounds for the struggle rating; participants' interpretations and stated reasoning are reported in Section~4.
\scriptsize
\Description{Attempt level table listing all 88 coded attempts grouped by participant, giving platform, task, elapsed time, completion, struggle, the basis for the struggle rating, clicks, scrolling, and interface states written as full screens S, popups P, and layovers L.}
\begin{longtable}{@{}p{0.5cm} p{1.0cm} p{1.6cm} p{1.0cm} p{1.5cm} p{0.6cm} p{4.2cm} p{0.6cm} p{0.6cm} p{0.8cm}@{}}
\caption{Detailed task performance by participant, platform, and task. S denotes full-screen transitions, P denotes popups, and L denotes layovers.}\\
\label{tab:participant-task-detail} \\
\toprule
\textbf{ID} & \textbf{Platform} & \textbf{Task} & \textbf{Time} & \textbf{Completion} & \textbf{Struggle} & \textbf{Basis for Rating} & \textbf{Clicks} & \textbf{Scrolls} & \textbf{Interface States} \\
\midrule
\endfirsthead
\multicolumn{10}{c}{\tablename\ \thetable\ -- \textit{Continued from previous page}} \\[4pt]
\toprule
\textbf{ID} & \textbf{Platform} & \textbf{Task} & \textbf{Time} & \textbf{Completion} & \textbf{Struggle} & \textbf{Basis for Rating} & \textbf{Clicks} & \textbf{Scrolls} & \textbf{Interface States} \\
\midrule
\endhead
\midrule
\multicolumn{10}{r}{\textit{Continued on next page}} \\
\endfoot
\bottomrule
\endlastfoot

\multirow{7}{*}{P1}
 & TikTok & Set Screen Time Limit & 1 min 52 sec & Completed independently & Low & Located screen time within settings and set a daily limit without redirection. & 16 & Normal & 9S+2P+2L \\[2pt]
 & TikTok & Make Account Private & 46 sec & Completed independently & Low & Navigated directly to account privacy and toggled the setting. & 7 & Normal & 6S \\[2pt]
 & TikTok & Manage In-App Notifications & 1 min 0 sec & Completed independently & Low & Entered notification settings and adjusted selected categories. & 9 & Normal & 4S \\[2pt]
 & TikTok & Report Content & 1 min 58 sec & Completed independently & Low & Brief initial pause, then located report through the three-dot menu. & 4 & Normal & 3S+2P+1L \\[2pt]
 & TikTok & Delete Account & 2 min 57 sec & Completed independently & Low & Located deactivate or delete under account information and completed the flow. & 12 & Normal & 9S+1P+2L \\[2pt]
 & Snapchat & Download My Data & 2 min 34 sec & Completed independently & Low & Reviewed data categories, selected a date range, and submitted the request. & 14 & Normal & 6S \\[2pt]
 & Snapchat & Delete Account & 1 min 55 sec & Completed independently & Low & Located delete account by scrolling within settings. & 4 & Normal & 5S+1L \\
\midrule

\multirow{6}{*}{P2}
 & YouTube & Set Screen Time Limit & 2 min 6 sec & Not completed & High & Searched settings without locating the control; the researcher ended the task. & 17 & Hesitation & 12S+1P+2L \\[2pt]
 & YouTube & Report Content & 1 min 18 sec & Completed independently & Low & Located report through the overflow menu. & 4 & Normal & 3S+1P \\[2pt]
 & Instagram & Set Screen Time Limit & 1 min 45 sec & Completed independently & Low & Located time management within settings and set a daily limit. & 10 & Normal & 5S \\[2pt]
 & Instagram & Make Account Private & 33 sec & Completed independently & Low & Located account privacy and toggled the setting. & 6 & Normal & 6S+1P \\[2pt]
 & Instagram & Report Content & 1 min 15 sec & Completed independently & Low & Located report and completed the reason-selection sequence. & 5 & Normal & 1S+5L \\[2pt]
 & Instagram & Delete Account & 3 min 53 sec & Completed independently & Moderate & Repeated backtracking through settings before locating deletion via in-settings search; no researcher guidance. & 34 & Hesitation & 10S+1P+15L \\
\midrule

\multirow{8}{*}{P3}
 & Snapchat & Manage In-App Notifications & 1 min 26 sec & Completed independently & Low & Navigated directly to notification settings and adjusted categories. & 4 & Normal & 4S \\[2pt]
 & Snapchat & Report Content & 1 min 16 sec & Completed independently & Low & Located report through the overflow menu and completed the flow. & 5 & Normal & 5S+1L \\[2pt]
 & Snapchat & Download My Data & 4 min 19 sec & Completed with assistance & High & Extended search across settings areas; researcher provided directional hints. & 35 & Normal & 13S+5P \\[2pt]
 & Snapchat & Delete Account & 2 min 8 sec & Completed independently & Low & Located delete account within settings without redirection. & 5 & Normal & 5S \\[2pt]
 & Instagram & Set Screen Time Limit & 1 min 34 sec & Completed independently & Low & Went directly to settings and located the limit control. & 4 & Normal & 3S \\[2pt]
 & Instagram & Make Account Private & 1 min 36 sec & Completed independently & Low & Account already private; setting confirmed from the privacy page. & 1 & Normal & 1S \\[2pt]
 & Instagram & Manage In-App Notifications & 1 min 06 sec & Completed independently & Low & Entered notification settings and reviewed the available categories. & 7 & Normal & 7S \\[2pt]
 & Instagram & Report Content & 40 sec & Completed with assistance & High & Researcher redirected the participant to the overflow menu. & 4 & Normal & 3S+1L \\
\midrule

\multirow{7}{*}{P4}
 & YouTube & Set Screen Time Limit & 3 min 31 sec & Completed independently & Moderate & Searched account and settings areas before locating the break reminder; recovered without guidance. & 8 & Normal & 7S+1P+11L \\[2pt]
 & YouTube & Manage In-App Notifications & 1 min 1 sec & Completed independently & Low & Located notification settings and confirmed the existing configuration. & 2 & None & 2S \\[2pt]
 & YouTube & Report Content & 10 min 50 sec & Completed with assistance & High & Entered the feedback flow first; researcher redirected to the in-video overflow menu. & 13 & Normal & 9S+2L \\[2pt]
 & TikTok & Set Screen Time Limit & 2 min 19 sec & Completed independently & Low & Located screen time in the activity centre and configured a daily limit and breaks. & 22 & Normal & 13S+1P+3L \\[2pt]
 & TikTok & Make Account Private & 45 sec & Completed independently & Low & Located the privacy page within settings and enabled private account. & 9 & Normal & 5S+2L \\[2pt]
 & TikTok & Report Content & 3 min 45 sec & Completed with assistance & High & Searched settings without recognising the long-press gesture; researcher prompted the gesture. & 11 & Normal & 7S+3L \\[2pt]
 & TikTok & Delete Account & 3 min 34 sec & Completed independently & Low & Located deactivate or delete under profile and completed the flow. & 11 & Normal & 10S+1P+2L \\
\midrule

\multirow{9}{*}{P5}
 & TikTok & Set Screen Time Limit & 1 min 54 sec & Completed independently & Low & Located screen time breaks and scheduled a reminder; confirmed placement before proceeding. & 12 & Normal & 9S+1P+3L \\[2pt]
 & TikTok & Make Account Private & 1 min 9 sec & Completed independently & Low & Located account privacy and toggled the setting. & 7 & Normal & 5S+2L \\[2pt]
 & TikTok & Manage In-App Notifications & 2 min 23 sec & Completed independently & Low & Entered notification settings and retained the existing configuration. & 2 & None & 2S \\[2pt]
 & TikTok & Report Content & 2 min 18 sec & Completed with assistance & High & Reached the long-press menu but did not identify report; researcher indicated the control. & 6 & Normal & 3S+2P+1L \\[2pt]
 & TikTok & Delete Account & 5 min 34 sec & Completed independently & Low & Located deactivate or delete independently and completed the flow through the data prompt. & 12 & Normal & 9S+1P+2L \\[2pt]
 & Instagram & Make Account Private & 1 min 38 sec & Completed independently & Low & Located account privacy under settings and toggled the setting. & 4 & Normal & 4S \\[2pt]
 & Instagram & Manage In-App Notifications & 2 min 35 sec & Completed independently & Low & Entered notification settings and reviewed the available options without extended search. & 3 & Normal & 3S \\[2pt]
 & Instagram & Report Content & 1 min 39 sec & Completed independently & Low & Located report through the overflow menu and completed reason selection. & 7 & Normal & 5S+1L \\[2pt]
 & Instagram & Delete Account & 2 min 12 sec & Completed independently & Low & Located account ownership and control and completed deletion. & 14 & Normal & 2S+1P+11L \\
\midrule

\multirow{9}{*}{P6}
 & Snapchat & Manage In-App Notifications & 3 min 41 sec & Completed independently & Low & Navigated directly to notifications and disabled selected categories. & 7 & Normal & 4S \\[2pt]
 & Snapchat & Download My Data & 3 min 51 sec & Completed with assistance & High & Searched profile and settings without locating the section; researcher provided a hint. & 12 & Hesitation & 10S \\[2pt]
 & Snapchat & Report Content & 3 min 28 sec & Completed with assistance & High & Searched the friends list rather than the content; researcher provided guidance before the report was submitted. & 24 & Normal & 10S \\[2pt]
 & Snapchat & Delete Account & 1 min 30 sec & Completed independently & Low & Located deactivate or delete under settings and selected permanent deletion. & 4 & Normal & 3S+1P \\[2pt]
 & TikTok & Set Screen Time Limit & 1 min 10 sec & Completed independently & Low & Located screen time immediately, then paused to distinguish the daily limit from screen time breaks before setting the limit. & 11 & Normal & 8S+3P \\[2pt]
 & TikTok & Make Account Private & 21 sec & Completed independently & Low & Located the privacy page within settings and enabled private account. & 2 & None & 2S \\[2pt]
 & TikTok & Manage In-App Notifications & 54 sec & Completed independently & Low & Returned to notifications and disabled selected update categories. & 6 & Normal & 2S \\[2pt]
 & TikTok & Report Content & 1 min 57 sec & Completed independently & Moderate & Used long press to reach the reporting tools; submitted twice and applied a restriction. & 10 & Normal & 4S+3L \\[2pt]
 & TikTok & Delete Account & 1 min 52 sec & Completed independently & Low & Located deactivate or delete and selected an alternative reason to avoid additional steps. & 10 & Normal & 5S+1P \\
\midrule

\multirow{9}{*}{P7}
 & TikTok & Set Screen Time Limit & 1 min 9 sec & Completed independently & Low & Prior user of the feature; set a limit without redirection. & 10 & Normal & 6S+2P \\[2pt]
 & TikTok & Make Account Private & 31 sec & Completed independently & Low & Located privacy within account settings and toggled the setting. & 4 & Normal & 2S+2P \\[2pt]
 & TikTok & Manage In-App Notifications & 1 min 17 sec & Completed independently & Low & Navigated notification settings and adjusted selected categories. & 9 & Normal & 3S+1P \\[2pt]
 & TikTok & Report Content & 5 min 4 sec & Completed independently & Low & Located report through the long-press menu and completed reason selection. & 5 & Normal & 4S+1L \\[2pt]
 & TikTok & Delete Account & 1 min 38 sec & Completed independently & Low & Located delete account and completed the flow. & 9 & Normal & 7S+1P+1L \\[2pt]
 & Instagram & Set Screen Time Limit & 1 min 7 sec & Completed independently & Low & Went directly to the control and set a daily limit; verified the setting before finishing. & 7 & Normal & 6S \\[2pt]
 & Instagram & Make Account Private & 44 sec & Completed independently & Low & Located account privacy and toggled the setting. & 3 & Normal & 2S \\[2pt]
 & Instagram & Report Content & 34 sec & Completed independently & Low & Completed the report flow through the overflow menu. & 5 & None & 3S+1L \\[2pt]
 & Instagram & Delete Account & 3 min 43 sec & Completed with assistance & High & Could not locate the deletion path; researcher directed the participant to accounts centre and personal details. & 22 & Hesitation & 20S \\
\midrule

\multirow{8}{*}{P8}
 & Instagram & Set Screen Time Limit & 1 min 10 sec & Completed independently & Low & Located the limit under settings and activity and set a daily limit. & 6 & Normal & 5S \\[2pt]
 & Instagram & Make Account Private & 31 sec & Completed independently & Low & Located account privacy and toggled the setting. & 3 & Normal & 1S+2L \\[2pt]
 & Instagram & Manage In-App Notifications & 29 sec & Completed independently & Low & Reached notification settings and retained all default categories. & 2 & Normal & 1S \\[2pt]
 & Instagram & Report Content & 1 min 33 sec & Completed independently & Low & Completed the report flow without redirection. & 8 & Normal & 2S+6L \\[2pt]
 & Instagram & Delete Account & 2 min 47 sec & Completed with assistance & High & Could not locate the deletion path; researcher provided step-by-step guidance. & 15 & Normal & 10S+3L \\[2pt]
 & TikTok & Set Screen Time Limit & 2 min 10 sec & Completed independently & Low & Prior user of the feature; located screen time via settings without redirection, carrying the route from the Instagram task. & 4 & Normal & 4S \\[2pt]
 & TikTok & Manage In-App Notifications & 32 sec & Completed independently & Low & Reached notification settings and made no changes. & 1 & None & 1S \\[2pt]
 & TikTok & Delete Account & 3 min 52 sec & Completed independently & Low & Located the deletion flow and completed it, skipping the pre-deletion questions. & 11 & Normal & 7S+1P \\
\midrule

\multirow{9}{*}{P9}
 & Snapchat & Manage In-App Notifications & 52 sec & Completed independently & Low & Navigated to notifications independently and enabled all categories. & 6 & Normal & 6S \\[2pt]
 & Snapchat & Report Content & 36 sec & Completed independently & Moderate & Could not begin until reportable content was available; then located report through long press. & 5 & Normal & 5S \\[2pt]
 & Snapchat & Download My Data & 1 min 43 sec & Completed independently & Low & Located my data by scrolling within settings. & 3 & Normal & 3S \\[2pt]
 & Snapchat & Delete Account & 2 min 8 sec & Completed independently & Low & Located delete account under settings without redirection. & 6 & Normal & 5S+1L \\[2pt]
 & TikTok & Set Screen Time Limit & 3 min 47 sec & Completed with assistance & High & Extended search within the settings area before requiring researcher guidance. & 10 & Hesitation & 7S+1P+1L \\[2pt]
 & TikTok & Make Account Private & 1 min 8 sec & Completed independently & Low & Located privacy and enabled private account immediately. & 4 & Normal & 2S+2P \\[2pt]
 & TikTok & Manage In-App Notifications & 3 min 25 sec & Completed independently & Low & Navigated to notifications and enabled all categories. & 10 & Normal & 4S+1P \\[2pt]
 & TikTok & Report Content & 56 sec & Completed independently & Low & Located the long-press entry without assistance, then traversed the multi-stage reason-selection sequence. & 7 & Normal & 3S+2L \\[2pt]
 & TikTok & Delete Account & 1 min 32 sec & Completed independently & Low & Located deactivate or delete and selected permanent deletion. & 9 & Normal & 7S+1P+1L \\
\midrule

\multirow{8}{*}{P10}
 & YouTube & Set Screen Time Limit & 4 min 46 sec & Completed with assistance & High & Searched privacy first; researcher reframed the label and indicated the control. & 21 & Hesitation & 20S+1L \\[2pt]
 & YouTube & Manage In-App Notifications & 55 sec & Completed independently & Low & Located notification settings and retained all categories enabled. & 3 & Normal & 3S \\[2pt]
 & YouTube & Report Content & 2 min 43 sec & Completed independently & Low & Located the overflow menu after initial uncertainty and completed the report. & 15 & Normal & 11S+4L \\[2pt]
 & Instagram & Set Screen Time Limit & 2 min 37 sec & Completed independently & Low & Located time management and set a limit. & 7 & Normal & 6S \\[2pt]
 & Instagram & Manage In-App Notifications & 1 min 19 sec & Completed independently & Low & Entered notification settings and worked through the available categories. & 7 & Hesitation & 7S \\[2pt]
 & Instagram & Report Content & 2 min 19 sec & Completed independently & Low & Located report through the overflow menu and completed the flow. & 3 & Normal & 1S+2P \\[2pt]
 & Instagram & Download My Data & 2 min 57 sec & Completed independently & Moderate & Extended search across settings before locating the control independently. & 14 & Hesitation & 13S+1L \\[2pt]
 & Instagram & Delete Account & 3 min 14 sec & Completed with assistance & High & Logged out rather than locating deletion; researcher redirected the participant. & 15 & Normal & 10S+3L \\
\midrule

\multirow{8}{*}{P11}
 & YouTube & Set Screen Time Limit & 1 min 33 sec & Completed independently & Low & Located time management while scrolling and set a bedtime and break reminder. & 6 & Hesitation & 2S+3P+1L \\[2pt]
 & YouTube & Manage In-App Notifications & 3 min 57 sec & Completed independently & Low & Located notifications and adjusted preferences across categories. & 7 & Hesitation & 1S \\[2pt]
 & YouTube & Report Content & 1 min 30 sec & Completed independently & Moderate & Located report in the overflow menu and completed the flow while uncertain about the outcome. & 6 & Normal & 1S+1P+2L \\[2pt]
 & Instagram & Set Screen Time Limit & 2 min 20 sec & Completed independently & Low & Located time management after expecting the YouTube structure; set a limit. & 14 & Normal & 4S+4L \\[2pt]
 & Instagram & Make Account Private & 34 sec & Completed independently & Low & Located account privacy within settings and toggled the setting. & 2 & Normal & 2S \\[2pt]
 & Instagram & Manage In-App Notifications & 1 min 43 sec & Completed independently & Low & Located notifications and moved through categories individually, deciding on each. & 4 & Normal & 2S+1L \\[2pt]
 & Instagram & Report Content & 3 min 23 sec & Not completed & High & Reached the content options menu through the documented gesture; no report option was present among the available options. & 11 & Hesitation & 8S+3P+3L \\[2pt]
 & Instagram & Delete Account & 5 min 34 sec & Completed with assistance & High & Searched settings and privacy areas without locating deletion; researcher provided step-by-step guidance. & 18 & Normal & 10S+3L \\

\end{longtable}
\normalsize

%
%

\end{document}